\documentclass[11pt]{article}
\usepackage{amsmath}
\usepackage{float}
\usepackage{hhline}
\usepackage{natbib}
\usepackage{color}
\usepackage{graphicx}
\usepackage{multirow}
\usepackage{bigints}
\usepackage{amssymb}
\usepackage[dvips]{epsfig}
\usepackage{color}
\usepackage{setspace}
\usepackage{soul,xcolor}
\usepackage{hyperref}
\usepackage{caption}
\usepackage{subcaption}

\newcommand{\key}[1]{\textcolor{red}{#1}}
\setstcolor{red}

\newcommand{\be}{\begin{eqnarray}}
\newcommand{\ee}{\end{eqnarray}}
\newcommand{\ba}{\begin{eqnarray*}}
	\newcommand{\ea}{\end{eqnarray*}}
\newtheorem{theorem0}{Theorem}
\newtheorem{lemma0}{Lemma}
\newtheorem{remark0}{Remark}
\newtheorem{fact0}{Fact}
\newtheorem{example0}{Example}
\newtheorem{corollary0}{Corollary}
\newtheorem{proposition0}{Proposition}
\newtheorem{conjecture0}{Conjecture}

\newenvironment{corollary}{\begin{corollary0} }{\end{corollary0}}

\def\boldfacefake #1{%
	\hbox{%
		\mathsurround=0pt
		\hbox to 0.4pt{$#1$\hss}%
		\hbox to 0.4pt{$#1$\hss}%
		\hbox {$#1$}%
	}%
}

\def\bftheta{\boldfacefake{\theta}}

\def\bftheta{\boldfacefake{\theta}}

\newcommand{\expect}{\mbox{\rm I\kern-.20em E}}
\newcommand{\reals}{\mbox{\rm I\kern-.20em R}}
\newcommand{\sreals}{\mbox{\small \rm I\kern-.20em R}}

\newcommand{\qed}{\nobreak \ifvmode \relax \else
      \ifdim\lastskip<1.5em \hskip-\lastskip
      \hskip1.5em plus0em minus0.5em \fi \nobreak
      \vrule height0.75em width0.5em depth0.25em\fi}

\newcommand{\ignore}[1]{}

\begin{document}
\title{\bf A simple consistent Bayes factor for testing the Kendall rank correlation coefficient}
\author{Shen Zhang, Keying Ye and Min Wang\thanks{Corresponding author: min.wang3@utsa.edu}\\
   Department of Management Science and Statistics\\
     The University of Texas at San Antonio, San Antonio, TX 78249-0634, USA}
\date{\today}        
\maketitle

\begin{abstract}
In this paper, we propose a simple and easy-to-implement Bayesian hypothesis test for the presence of an association, described by Kendall's $\tau$ coefficient, between two variables measured on at least an ordinal scale. Owing to the absence of the likelihood functions for the data, we employ the asymptotic sampling distributions of the test statistic as the working likelihoods and then specify a truncated normal prior distribution on the noncentrality parameter of the alternative hypothesis, which results in the Bayes factor available in closed form in terms of the cumulative distribution function of the standard normal distribution. Investigating the asymptotic behavior of the Bayes factor we find the conditions of the priors so that it is consistent to whichever the hypothesis is true. Simulation studies and a real-data application are used to illustrate the effectiveness of the proposed Bayes factor. It deserves mentioning that the proposed method can be easily covered in undergraduate and graduate courses in nonparametric statistics with an emphasis on students' Bayesian thinking for data analysis.


\end{abstract}

\newpage

\section{Introduction} \label{Section:01}

The Kendall rank correlation coefficient, often referred to as Kendall's $\tau$ coefficient, is one of the commonly used nonparametric statistics to assess the statistical dependence between two variables measured on at least an ordinal scale (\citeauthor{Kend:1938} \citeyear{Kend:1938}). It is well-known that Kendall's $\tau$ coefficient is carried out on the ranks of the observed data, and thus, it is generally  resistant to outliers and tolerates violations to its normality assumption of the data. Furthermore, it is invariant under rank-preserving transformation on the scales of measurement; see, for example, \cite{Krus:1958, Kenn:Gibb:1990, Wass:2006}, among others.

Let $(x_i, y_i)$ be the $i$th observation of the joint random variables $X$ and $Y$ for $i = 1, \cdots, n$. Let $\tau \in [-1, 1]$ stand for the Kendall's population correlation, which measures an ordinal association between $X$ and $Y$. We are interested in testing the hypotheses of the form
\begin{equation} \label{test:setting:01}
H_0: \tau =  \tau_0 \ \ \mathrm{versus} \ \ H_1: \tau \neq  \tau_0,
\end{equation}
where $\tau_0 \in [-1, 1]$ is a known value. Within a frequentist framework,
researcher usually adopt a modified version of Kendall's $\tau$ correlation for the hypothesis testing problem (e.g., \citeauthor{samara1988test}, \citeyear{samara1988test}). One of the modified versions is the standardized Kendall's $\tau$ correlation defined as
\begin{equation} \label{T:test}
   T^\ast =\frac{\sum\limits_{1\leq i< j\leq n}Q((x_i,y_i),(x_j, y_j)) - \tau_0}{\sqrt{n(n-1)(2n+5)/18}},
\end{equation}
where $Q(\cdot, \cdot)$ is the concordance indicator function given by
\begin{equation*}
    Q\bigl((x_i,y_i),(x_j, y_j) \bigr)=\left\{\begin{array}{cc}
       -1  & ~\text{if}~(x_i-x_j)(y_i - y_j)<0,  \\
        1 &  ~\text{if}~(x_i-x_j)(y_i - y_j)>0.
    \end{array}\right.
\end{equation*}
It is known from \cite{Noet:1955} that under $H_0$, $T^\ast$ approximately follows a standard normal distribution. Thus, we can estimate the p-value $p = 2P(Z >|T^\ast|)$, where $Z$ stands for the standard normal distribution. At the $\alpha$ significance level, we could reject $H_0$ if $p < \alpha$.

Bayesian analysis has been widely adopted as an alternative to the frequentist procedures for hypothesis testing problems in recent years, partly due to computational advances but also because of its practical advantages in interpretability. Within the Bayesian framework, a natural way for comparing two competing models under the hypotheses is to use the Bayes factor (\citeauthor{Kass:95} \citeyear{Kass:95}), which is defined as the ratio of the marginal likelihood functions under the two hypotheses and is given by
\begin{equation} \label{BF:00}
\mathrm{BF}_{01}=\frac{P(\text{data} \mid H_0)}{P(\text{data} \mid H_1)},
\end{equation}
where $P(\text{data} \mid H_j)$ is the marginal posterior density of the data under $H_j$ and is given by
\begin{equation*}
P(\text{data} \mid H_j)=\int P(\text{data} \mid \bftheta_j, H_j)\pi_j(\bftheta_j \mid  H_j)\, d\bftheta_j,
\end{equation*}
with $P(\text{data} \mid \text{\boldmath{$\theta$}}_j, H_j)$ being the sampling density of the data with the unknown parameter $\bftheta_j$ and $\pi_j(\cdot)$ being the specified prior for $\bftheta_j$ under $H_j$ for $j=0, 1$. It deserves mentioning that the Bayes factor in (\ref{BF:00}) can be viewed as a weighted average likelihood ratio that provides the relative plausibility of the data under the two competing hypotheses. For instance, $\mathrm{BF}_{01} = 10$ indicates that the data are 10 times more likely to be generated from the model under $H_0$ than under $H_1$. \cite{Jeff:1961} and \cite{Kass:95} provided the classification categories of evidential strength for interpreting the Bayes factor based on the data. For making a dichotomous decision, we may reject $H_0$ when the $\mathrm{BF}_{01}$ is smaller than a certain specified threshold (e.g. $\mathrm{BF}_{01} < 1$).

In order to derive the Bayes factor in (\ref{BF:00}), one needs to have a working likelihood function for the data under each hypothesis and then specify the prior distributions for the unknown parameters of each likelihood function. However, owing to the absence of the likelihood function of the data in the nonparametric settings, development of the Bayes factor for testing $\tau$ in (\ref{test:setting:01}) has received relatively little attention in the literature. To address this difficulty, \cite{Yuan:John:2008} first reformulated the alternative hypothesis with the Pitman translation alternative (\citeauthor{Rank:Wolf:1979} \citeyear{Rank:Wolf:1979}) and then introduced a noncentrality parameter to distinguish between $H_0$ and $H_1$ in (\ref{test:setting:01}). Thereafter, instead of working on the sampling distributions of the data, they employed the asymptotic sampling distributions of the test statistics in (\ref{T:test}) and derived the Bayes factor based on the normal distribution prior on the noncentrality parameter. Since the resulting Bayes factor depends on a hyperparameter that controls the expected departure from the null hypothesis, they adopted the data-dependent maximization for this hyperparameter and obtained the upper bound of the resulting Bayes factor in favor of the null hypothesis. Meanwhile, they also pointed out that this upper bound may not be practically useful unless some constraints are imposed for $\tau$.
To deal with this practical implementation issue of the Bayes factor related to this hyperparameter, \cite{Doorn:Ly:Wage:2018} recently  eliminated the choice of this hyperparameter based on a `parametric yoking' procedure by matching a prior for $\tau$ with its parametric ones: they transformed the beta prior distribution for the Pearson correlation coefficient $\rho$ through Greiner’s relation between $\tau$ and  $\rho$ (i.e., $\tau = 2/\pi\mathrm{arcsin}(\rho)$) for bivariate normal data (\citeauthor{Krus:1958} \citeyear{Krus:1958}). However, such a relationship may not be valid when the data are not bivariate normal. This observation motivates us to review the practical issues of the Bayes factor for the Kendall's $\tau$ correlation.

In this paper, we suggest the direct use of the truncated normal distribution as an alternative prior for $\tau$, instead of using the parametric yoking procedure by \cite{Doorn:Ly:Wage:2018}. Even though there exist a variety of other possible priors, we justify that the truncated normal prior not only allows researchers to directly incorporate available prior information into Bayesian data analysis. The proposed Bayesian methodology could improve and enrich the literature in the following aspects:
\begin{itemize}
    \item The proposed Bayes factor can be used for testing the hypotheses of the form in (\ref{test:setting:01}) for any $\tau_0 \in [-1, 1]$.
    \item The proposed Bayes factor depends on the data only through the statistic in (\ref{T:test}) and has a simple closed-form expression and can thus be easily calculated by practitioners, so long as they are familiar with the Kendall's $\tau$ coefficient.
    \item The proposed Bayes factor is consistent whichever the true hypothesis in (\ref{test:setting:01}) is under certain choices of prior parameters.
    \item Our result can be easily covered in undergraduate and graduate courses on nonparametric statistics with an emphasis on students' Bayesian thinking about nonparametric hypothesis testing to real-data problems.
\end{itemize}

The remainder of this paper is organized as follows. In Section \ref{section:02}, we review the Bayesian formulation of Kendall's $\tau$ pioneered by \cite{Yuan:John:2008} and derive the Bayes factor under the truncated normal prior. The consistency property of the resulting Bayes factor is also examined in this section. In Section \ref{section:03}, we conduct simulation studies to compare the finite sample performance of the proposed Bayes factor with several existing ones in the literature. A real application is provided in Section \ref{section:04} for illustrative purposes. Finally, some concluding remarks are provided in Section \ref{section:05} with proofs of the main results deferred to the Appendix.

\section{Bayesian inference} \label{section:02}

In Section \ref{section:02:01}, we review the Bayesian framework for testing the presence of Kendall's $\tau$ correlation from \cite{Yuan:John:2008} and \cite{Doorn:Ly:Wage:2018}. We develop the Bayes factor based on the truncated normal prior in Section \ref{section:02:02} and elaborate the selection of the prior hyperparameters in a specific context in Section \ref{section:02:03}.

\subsection{Bayesian modeling of test statistics} \label{section:02:01}

In nonparametric hypothesis testing framework, the sampling distributions of the data are usually not specified, resulting in difficulties in defining  Bayesian hypothesis test. To deal with this issue, \cite{Yuan:John:2008}
adopted the results from the  asymptotic theory of $U$-statistics and linear rank statistics to derive the asymptotic distribution of the test statistic under the alternative hypothesis by using the Pitman translation alternative. To be more specific,  \cite{Yuan:John:2008} took the form of Pitman translation alternative for the alternative hypothesis in (\ref{test:setting:01}) and proposed the following hypotheses
\begin{equation} \label{old:pitman}
H_0: \tau = \tau_0 \ \ \mathrm{versus} \ \ H_1: \tau = \tau_0 +  \frac{\Delta}{\sqrt{n}},
\end{equation}
where $\Delta \in \bigl[-\sqrt{n}(1 + \tau_0), \sqrt{n}(1  - \tau_0)\bigr]$ stands for a bounded noncentrality parameter to distinguish between the null and alternative hypotheses. In so doing, \cite{Yuan:John:2008} showed that under certain assumptions (e.g., see Lemma 1 of  \citeauthor{Yuan:John:2008},  \citeyear{Yuan:John:2008}), the sampling distribution of the test statistic $T^\ast$ follows a limiting normal distribution with mean $C\Delta$ and variance 1 under $H_1$,  where $C$ is the asymptotic efficacy parameter of the test and is equal to 3/2 for Kendall's $\tau$ test statistic $T^\ast$. It deserves mentioning that the asymptotic distribution of $T^\ast$ under $H_1$ depends only on an unknown scaling parameter in which a prior density can be specified.

It is known that under $H_0$, $T^\ast$ follows a limiting standard normal distribution, denoted by $T^\ast \sim \text{N}(0, 1)$, Thus, \cite{Yuan:John:2008} adopted these known asymptotic distributions of $T^\ast$ under the null and alternative hypotheses as the working likelihood functions required for the definition of the Bayes factor in (\ref{BF:00}) by specifying a normal prior for the unknown scaling parameter $\Delta$, such that $\Delta \sim N(0, \delta_n^2)$, where the hyperparameter $\delta_n$ represents the expected departure from the null value of $\Delta$. The resulting Bayes factor for testing the transformed hypotheses in (\ref{old:pitman}) is given by
\begin{equation} \label{previous:BF}
\mathrm{BF}_{01} = \sqrt{1+\frac{9}{4}\delta_n^2} ~\exp \biggl(-\frac{\delta_n^2 T^{*2}}{2\delta_n^2+8/9}\biggr),
\end{equation}
which has a simple closed-form expression that depends on the data through $T^{*2}$. In practice, a relatively large value of $\delta_n$ may be preferred to minimize the prior information. However, when $\delta_n$ becomes sufficiently large, $\mathrm{BF}_{01}$ in (\ref{previous:BF}) approaches infinity, indicating that it favors the null hypothesis, regardless of the data information. This phenomenon is called Bartlett’s paradox, see, for example, \cite{Bart:1957, Jeff:1961}. To tackle this issue, \cite{Yuan:John:2008} adopted the maximum marginal likelihood estimator for this hyperparameter, which results in an upper bound of the Bayes factor in favor of the null hypothesis. In the meantime, they also mentioned that this upper bound may not be practically useful as evidence in decision-making unless some constraints are imposed for the choice of $\delta_n$.

Recently, \cite{Doorn:Ly:Wage:2018} pointed out that the Pitman translation alternative relies on the sample size and it may thus become indistinguishable to the null hypothesis as the sample size tends to infinity. To avoid this potential issue, they redefined the hypotheses in (\ref{old:pitman}) as
\begin{equation} \label{test:hyp}
H_0: \tau = \tau_0 \ \ \mathrm{versus} \ \ H_1: \tau = \tau_0 +  \Delta_n,
\end{equation}
where $\Delta_n \in [-1 - \tau_0, 1 - \tau_0]$ becomes synonymous with the true value $\tau$ when $\tau_0 = 0$, Thereafter, \cite{Doorn:Ly:Wage:2018} specified a parametric yoking prior on $\Delta_n$, which is obtained by using Greiner’s relation between the Kendall's $\tau$ and the Pearson correlation for bivariate normal data, whereas such a relationship may not be valid when the data are not bivariate normal. This observation motivates us to consider some alternative priors for $\Delta_n$ on the domain $[ -1-\tau_0, ~1-\tau_0]$. Specifically, in this paper, we recommend the use of  the truncated normal prior distribution, since it not only allows researchers to incorporate available prior information for making posterior inference, but also keeps a closed-form Bayes factor in terms of the cdf of a standard normal distribution discussed in the following section.

\subsection{The Bayes factor based on the truncated normal prior} \label{section:02:02}

Suppose that $\mathcal{T}$ is a normally distributed random variable with mean $\mu$ and variance $\sigma^2$ and lies within the interval $[c_1, c_2]$ with $-\infty < c_1 < c_2 < \infty$, denoted by $\mathcal{T} \sim TN(\mu, \sigma^2, c_1, c_2)$. The probability density function (pdf) for $\mathcal{T}$ is given by
\begin{equation*}
f(t; \mu,\sigma^2, c_1, c_2)=\frac{1}{\sigma}{\frac {\phi \bigl({\frac {t-\mu}{\sigma}}\bigr)}{ \Phi \bigl({\frac {c_2- \mu }{\sigma }}\bigr)-\Phi \bigl({\frac {c_1-\mu }{\sigma }}\bigr)}},\text{ for }t\in[c_1, c_2],
\end{equation*}
where $\phi( \cdot)$ and $\Phi (\cdot)$ stand for the pdf and cdf of the standard normal distribution, respectively. It can be seen from Figure \ref{FIG:equal} that the truncated normal distribution is very flexible for modeling different types of data and thus allows researchers to incorporate available prior information for drawing Bayesian inference. Using the truncated normal prior, we present the Bayes factor in the following theorem.

\begin{theorem0} \label{theorem:01}
For the hypotheses in (\ref{test:hyp}), the Bayes factor in (\ref{BF:00}) under the truncated normal prior $\Delta_n \sim TN(\lambda_n, \kappa_n^2, -1 - \tau_0, 1 - \tau_0)$ turns out to be
\begin{equation} \label{PBF}
\mathrm{BF}_{01} =\sqrt{\frac{9n\kappa_n^2}{4}+1} ~\exp \biggl\{\frac{1}{2}\Bigl(\frac{\lambda_n^2}{\kappa_n^2}- \frac{\mu_n^2 }{\sigma_n^2} \Bigr)  \biggr\}\cdot \frac{\Phi\bigl(\frac{1-\tau_0-\mu_n}{\sigma_n}\bigr)-\Phi\bigl(\frac{-1-\tau_0-\mu_n}{\sigma_n}\bigr)}{\Phi\bigl(\frac{1-\tau_0-\lambda_n}{\kappa_n}\bigr)-\Phi\bigl(\frac{-1-\tau_0-\lambda_n}{\kappa_n}\bigr)},
\end{equation}
where
\begin{equation}\label{eq:postMeanVar}
\mu_n=\frac{\frac{T^\ast 3\sqrt{n}}{2}+ \frac{\lambda_n}{\kappa_n^2}}{\frac{9n}{4} + \frac{1}{\kappa_n^2} }
\quad \mathrm{and} \quad \sigma_n = \sqrt{\frac{1}{\frac{9n}{4} + \frac{1}{\kappa_n^2}}}.
\end{equation}
\end{theorem0}

The proof of Theorem \ref{theorem:01} is left in the Appendix. We observe that the Bayes factor in (\ref{PBF}) has a simple closed-form expression and can be easily calculated by practitioners with an Excel spreadsheet, so long as they are familiar with the Kendall's $\tau$ coefficient. In addition, analogous to the Bayes factor in (\ref{previous:BF}) by \cite{Yuan:John:2008}, it relies on the data only through $T^\ast$ in (\ref{T:test}) and can be regarded as a new Bayesian version of Kendall's $\tau$ coefficient.
{\begin{figure}[!htbp]
\centering \scalebox{0.85}{\includegraphics{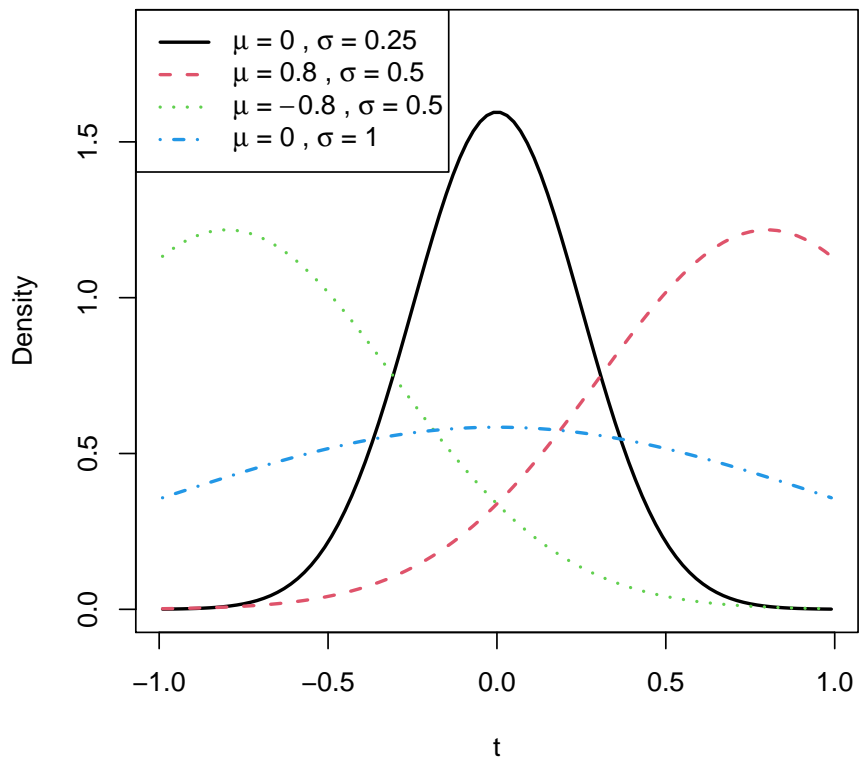}}
\caption{The pdf graphs of the truncated normal distribution with $c_1 = -1$, $c_2 = 1$, and different values of $\mu$ and $\sigma$.} \label{FIG:equal}
\end{figure}
}

From a Bayesian theoretical viewpoint, it is of interest to investigate the consistency of the Bayes factor when the sample size approaches infinity. In general, consistency means that the true hypothesis will be chosen if enough data are provided, assuming that one of them is true (e.g., \citeauthor{Fern:2001} \citeyear{Fern:2001}). The following theorem justifies the consistency of the proposed Bayes factor for the hypotheses in (\ref{test:hyp}) with its proof deferred to the Appendix.

\begin{theorem0} \label{theorem:02}
For testing the hypotheses in  (\ref{test:hyp}), we assume that the prior distribution of
$$
\Delta_n \sim TN\bigl(\lambda_n, \kappa_n^2, -1 - \tau_0, 1 - \tau_0\bigr), \text{ with }\lambda_n = O(n^{-a})\text{ and } \kappa_n = O(n^{-b}).
$$
As the sample size tends to infinity, the Bayes factor in (\ref{PBF}) for the hypothesis testing problem in  (\ref{test:hyp}) is consistent whichever hypothesis is true when $0\le a\le b<{1}/{2}$.
\end{theorem0}
We observe from this theorem that the asymptotic behavior of the Bayes factor relies on the rates of the two hyperparameters with respect to the sample size, in which the decreasing rate of $\lambda_n$ cannot be faster than the one of $\kappa_n$. In other words, the restriction on the hyperparameter spaces such that $0 \leq a \leq b < 1/2$ is necessary for guaranteeing the consistency of the Bayes factor under the null hypothesis. Note that the situation considered in \cite{Yuan:John:2008} that $\Delta_n = {\Delta}/{\sqrt{n}}$ where $\Delta$ is assumed to be a normal distribution with mean 0 and variance $\delta_n^2$ will not lead to the consistency of the Bayes factor in (\ref{previous:BF}) unless $\delta_n = O(n^{1/2})$.

Furthermore, in the absence of prior information for $\Delta_n$ in (\ref{test:hyp}), it is quite reasonable to consider a truncated normal prior distribution with the two hyperparameters independent of the sample size (e.g., $\lambda_n =0$ and $\kappa_n = 1$). In such a case, Theorem \ref{theorem:02} can be directly applied, with $a = b = 0$, to show the consistency of the Bayes factor summarized in the following corollary.

\begin{corollary}\label{cor}
When we assume that $\Delta_n\sim TN(\mu_0,\sigma_0,-1-\tau_0,1-\tau_0)$, where $\mu_0$ and $\sigma_0$ are known constants independent of the sample size, the Bayes factor in (\ref{PBF}) is consistent whichever hypothesis is true.
\end{corollary}

\ignore{For the cases considered in literature, for instance, \cite{Yuan:John:2008}, they assume $\Delta_n =\frac{\Delta}{\sqrt{n}}$ with $\Delta$ following a standard normal distribution. This is a case when $\lambda_n =0$ and \key{need more descriptions}.}

\subsection{The selection of the hyperparameters} \label{section:02:03}

At this stage, we need to discuss the specification of the hyperparameters $\lambda_n$ and $\kappa_n$ of the truncated normal prior for $\Delta_n$ in Theorem \ref{theorem:01}, which plays an important role in determining the performance of the Bayes factor in (\ref{PBF}). When no prior information is available, it is natural to specify $\lambda_n = 0$ and a relatively large value for $\kappa_n$ (e.g., $\kappa_n = 2$), since the prior becomes flatter as $\kappa_n $ increases, resulting in the prior with weak prior information. When prior information is available, it seems beneficial to incorporate available information into the Bayes factor with respect to an appropriate selection of the two hyperparameters.

\begin{figure}[!h]
\centering
 \begin{subfigure}[b]{0.48\textwidth}
         \centering
\includegraphics[width=81mm]{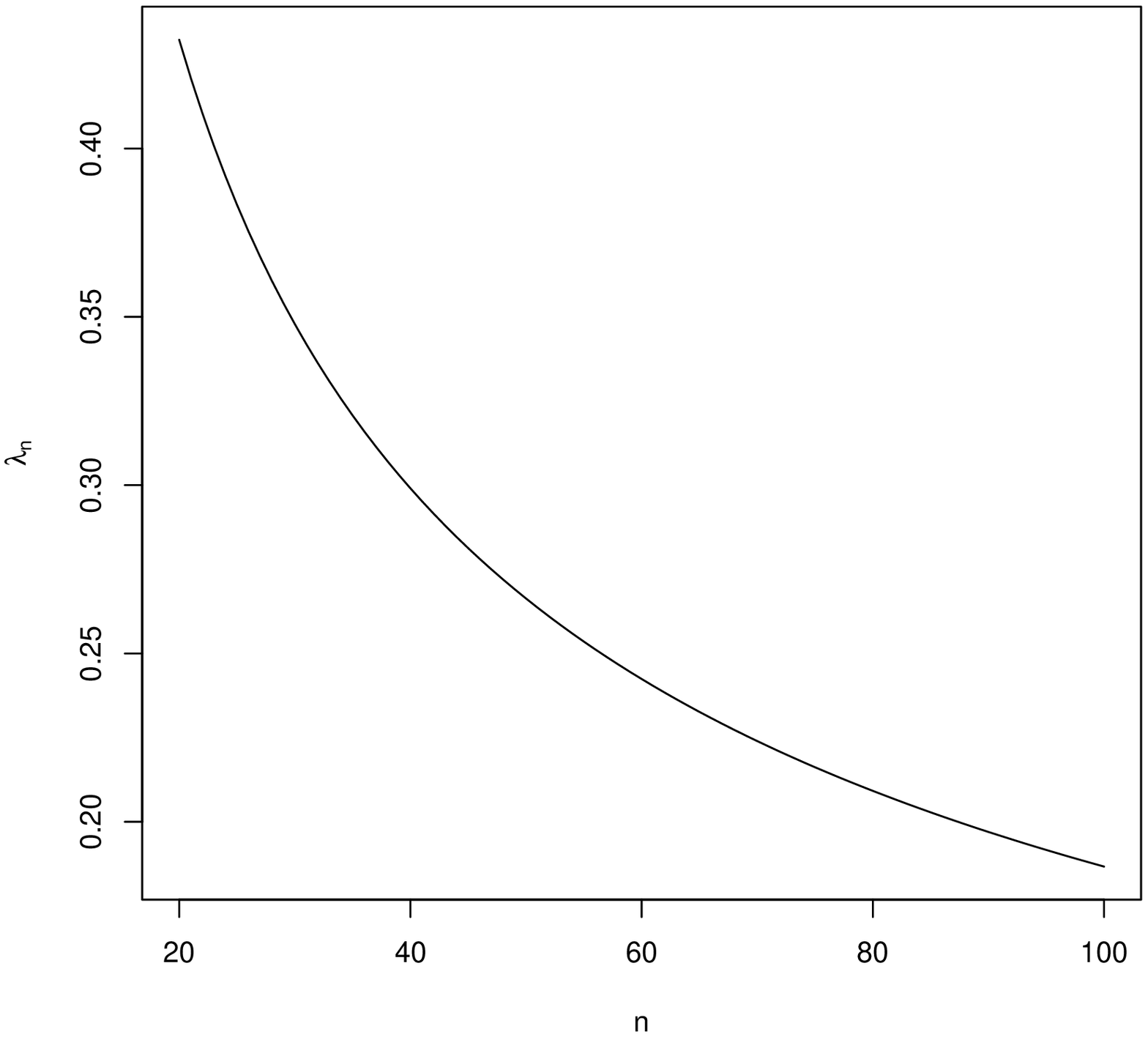}
\caption{}
    \label{figa}
     \end{subfigure}
\begin{subfigure}[b]{0.48\textwidth}
\includegraphics[width=81mm]{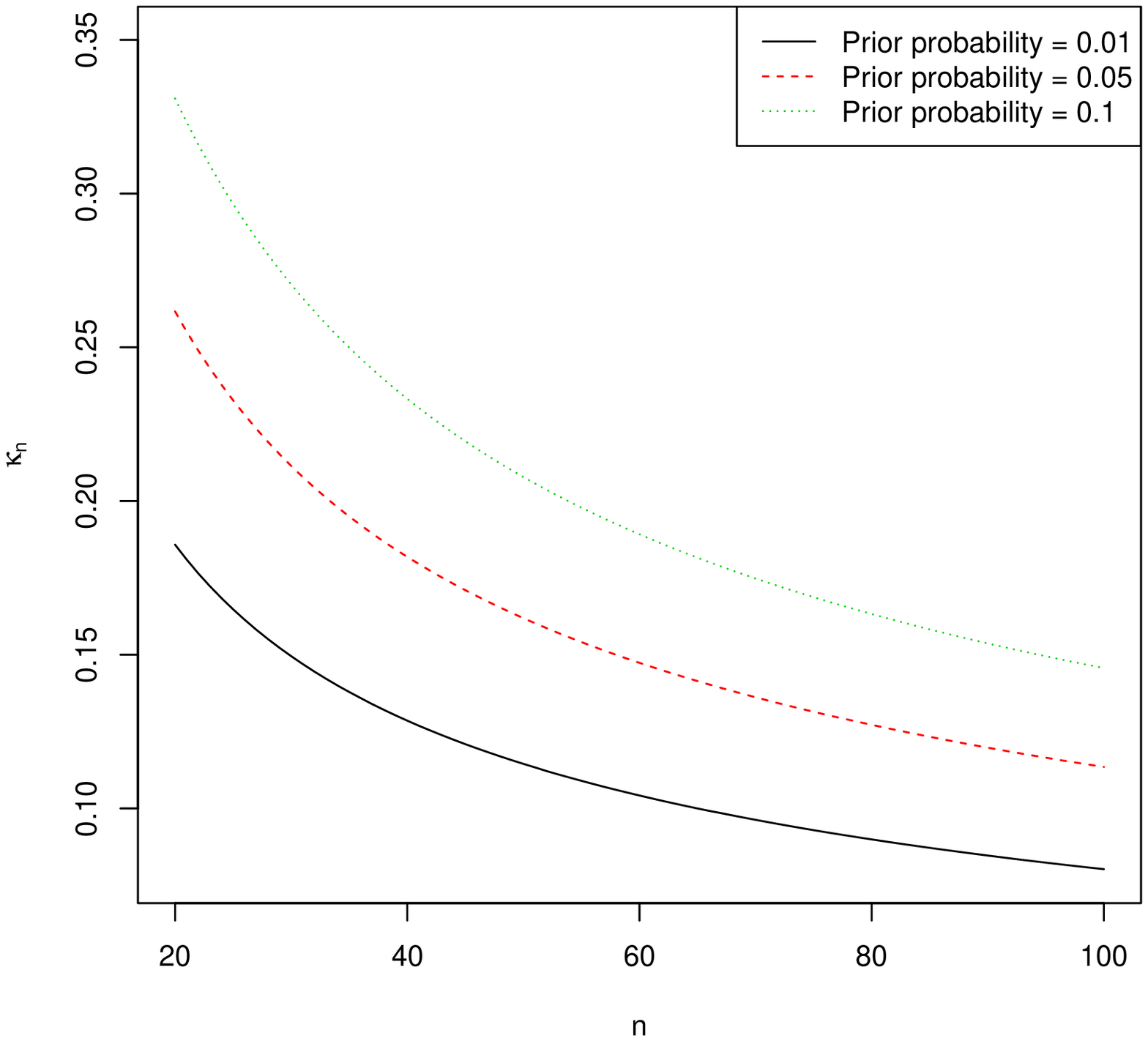}
\caption{}
\label{figb}
 \end{subfigure}
\caption{The hyperparameters $\lambda_n$ and $\kappa_n^2$  of the truncated normal prior as a function of $n$.} \label{FIG:priormean}
\end{figure}

As an illustration, it is known that prior information about the expected effect size (i.e., the value of $\lambda_n$) is usually used for the sample size calculation in experimental research designs. For a two-tailed test of the null hypothesis in (\ref{test:setting:01}), let $\tau_1$ be the alternative value of the Kendall's $\tau$ coefficient that we wish to detect. The required sample size (\citeauthor{Bone:Wrig:2000} \citeyear{Bone:Wrig:2000}) for detecting the value of $\tau_1$ with power $1-\beta$ and a significance level of $\alpha$ is given by
$$
n = 4 + 0.437\biggl(\frac{Z_{\alpha/2} + Z_{\beta}}{Z(\tau_1) - Z(\tau_0)}\biggr)^2,
$$
where $\beta$ stands for the Type II error probability, $Z_\xi$ is the upper $\xi$-percentage point of the standard normal distribution, and $Z(\tau_{b}) $ is the Fisher's $z$-transformation given by $Z(\tau_{b}) = 0.5\ln\bigr[(1+ \tau_b)/(1 -\tau_b)\bigr]$. It can be easily shown that
$$
\tau_1 = \frac{\exp\left\{{2(Z_{\alpha/2} + Z_{\beta})}/{\sqrt{\frac{n-4}{0.437}}}\right\}- \frac{1 - \tau_{0}}{1 + \tau_{0}}}{\exp\left\{{2(Z_{\alpha/2} + Z_{\beta})}/{\sqrt{\frac{n-4}{0.437}}}\right\} + \frac{1 - \tau_{0}}{1 + \tau_{0}}}.
$$
We here consider a case study to illustrate how the two hyperparameters of the truncated normal prior can be specified. Suppose that we are interested in testing the presence of the Kendall's correlation between the two random variables (i.e., $\tau_0 = 0$). In this study with powered at $80\% = 100(1-\beta)\%$ and $\alpha = 0.05$, the expected standardized effect size is given by
$$
\tau_1 = \frac{\exp\left\{5.6/\sqrt{(n-4)/0.437}\right\}- 1}{\exp\left\{5.6/\sqrt{(n-4)/0.437}\right\} + 1},
$$
which may be treated as the mean of the truncated normal prior, such that $\lambda_n = \tau_1$. Under this setup, if $n = 50$, then we may choose $\lambda_n = 0.266$, which corresponds to the medium Cohen (\citeauthor{Cohen:1988}, \citeyear{Cohen:1988}) classified effect size defined by \cite{botsch2011chapter}.
More generally, as shown in Figure \ref{figa},  the prior mean $\lambda_n$ is a decreasing function of the sample size $n$.

For the choice of $\kappa_n$, we observe that the value of $\kappa_n$ can be expressed as a function of the prior probability such that the effect size is expected to be in the wrong direction. For instance, when we expect that the prior probability $P(\Delta < 0 \mid \Delta \neq 0)  = 0.1$ with  $n=50$,  we may choose $\lambda_n = 0.266$ and $\kappa_n = 0.207545$ to achieve such an expectation. We observe from Figure \ref{figb} that the hyperparameter $\kappa_n$ is also a decreasing function of $n$ with various choices of the prior probabilities and that the value of $\kappa_n$ increases with the prior probability when $n$ is fixed. It deserves mentioning that in practical applications, we should consider other values of these hyperparameters to ensure the consistency of the results.

\section{Simulation studies} \label{section:03}

In this section, we carry out simulation studies to investigate the finite sample performance of the Bayes factor in (\ref{PBF}), denoted by BF$_\mathrm{tnorm}$ for short. Specifically, we conduct simulation studies to assess its sensitivity with respect to the hyperparameters $(\lambda_n, \kappa_n)$ in Section \ref{simulation:01} and compare its performance with some existing testing procedures in Section \ref{simulation:02}. The decision criterion is to reject $H_0$ if the value of the Bayes factor BF$_{01}$ is smaller than 1 and $H_1$, otherwise.

\subsection{Simulation 1} \label{simulation:01}

We simulate bivariate data for a given value of Kendall's $\tau$ as follows. Due the the absence of the likelihood function of the data, we follow the idea of \cite{Doorn:Ly:Wage:2018} by using the copula, which are multivariate distribution functions whose one-dimensional margin distributions are uniform on the interval $[0, 1]$. In our simulation studies, we considered four types of copulas: the Clayton, Frank, Gumbel, and normal copulas and reached similar conclusions. Thus, in this paper, we only focus attention on the normal copula by utilizing Greiner's relation (\citeauthor{Greiner:1909} \citeyear{Greiner:1909}) such that $\tau = 2 \arcsin(\rho)/ \pi$, where $\rho$ stands for the Pearson correlation coefficient. To generate bivariate data based on the Kendall's $\tau$, we simulate $1,000$ random samples of size $n$ from the 2-dimensional Gaussian copula with correlation matrix $\Sigma = \begin{pmatrix}
1 & \rho\\
\rho & 1
\end{pmatrix}$, whose cdf is given by
$$
c_\Sigma(x, y) = \Phi_\Sigma \left( \Phi^{-1}(x) , \Phi^{-1}(y)\right),
$$
where $\Phi_\Sigma$ is the joint bivariate distribution function of a normal variable with mean vector zero and correlation matrix $\Sigma$ and $\Phi^{-1}$ is the inverse of the standard normal cdf.

In the simulation study, we consider $n =\{10, 30, 50, 100\}$ to investigate the impact of the sample size to the Bayes factor. For each sample size, we generate $10,000$ random samples from the Gaussian copula with the value of $\tau$ ranging from $-1$ to 1 in increments of 0.1. To assess the sensitivity of the hyperparameters, we take $\kappa_n = \{0.25, 0.5, 1, 2\}$, since it seems natural to specify $\lambda_n = 0$ to represents our lack of information about the true parameter $\tau$. The decision criterion used in this paper is to choose $H_1$ if the Bayes factor BF$_\mathrm{tnorm} > 1$ and $H_0$ otherwise.

The relative frequencies of rejecting $H_0$ based on BF$_\mathrm{tnorm}$ with different choices of $n$ and $\kappa_n$ are depicted in Figure \ref{FIG:FR2}.
We observe that when the sample size is small, BF$_\mathrm{tnorm}$ is somehow sensitive to the small values of $\kappa_n$, such as $\kappa_n =0.25$ and $n = 10$. However, as $\kappa_n$ increases to 1 or above, BF$_\mathrm{tnorm}$ behaves similarly, even when the sample size is small. For instance, when $n = 50$, BF$_\mathrm{tnorm}$ with $\kappa_n = \{1, 2\}$ are almost identical in terms of the relative frequency of rejection of $H_0$. In addition, as suggested by the reviewer, it deserves pointing out that when $\kappa_n$ is small, the considered prior under the alternative hypothesis concentrates more on the case of $\tau = 0$, indicating that both the null and alternative hypotheses tend to become less discriminative. Consequently, we expect that when $\kappa_n$ is small, the proposed Bayes factor BF$_\mathrm{tnorm}$ results in more rejections of $H_0$, as shown in Figure \ref{FIG:FR2}. A similar behavior of the Bayes factor  has also been observed for comparing two-sample population means by \cite{Wang:Liu:2016}. Consequently, in the absence of prior knowledge, we do not suggest the use of small values of $\kappa_n$ to ensure the reliable results of the Bayes factor for quantifying the evidence in favor of the null and alternative hypotheses, respectively. Numerical results from simulation studies show that we advocate the value of $\kappa \geq 1$ to minimize prior inputs for the Bayes factor.

{
\begin{figure}[!h]
\centering
\includegraphics[width=80mm]{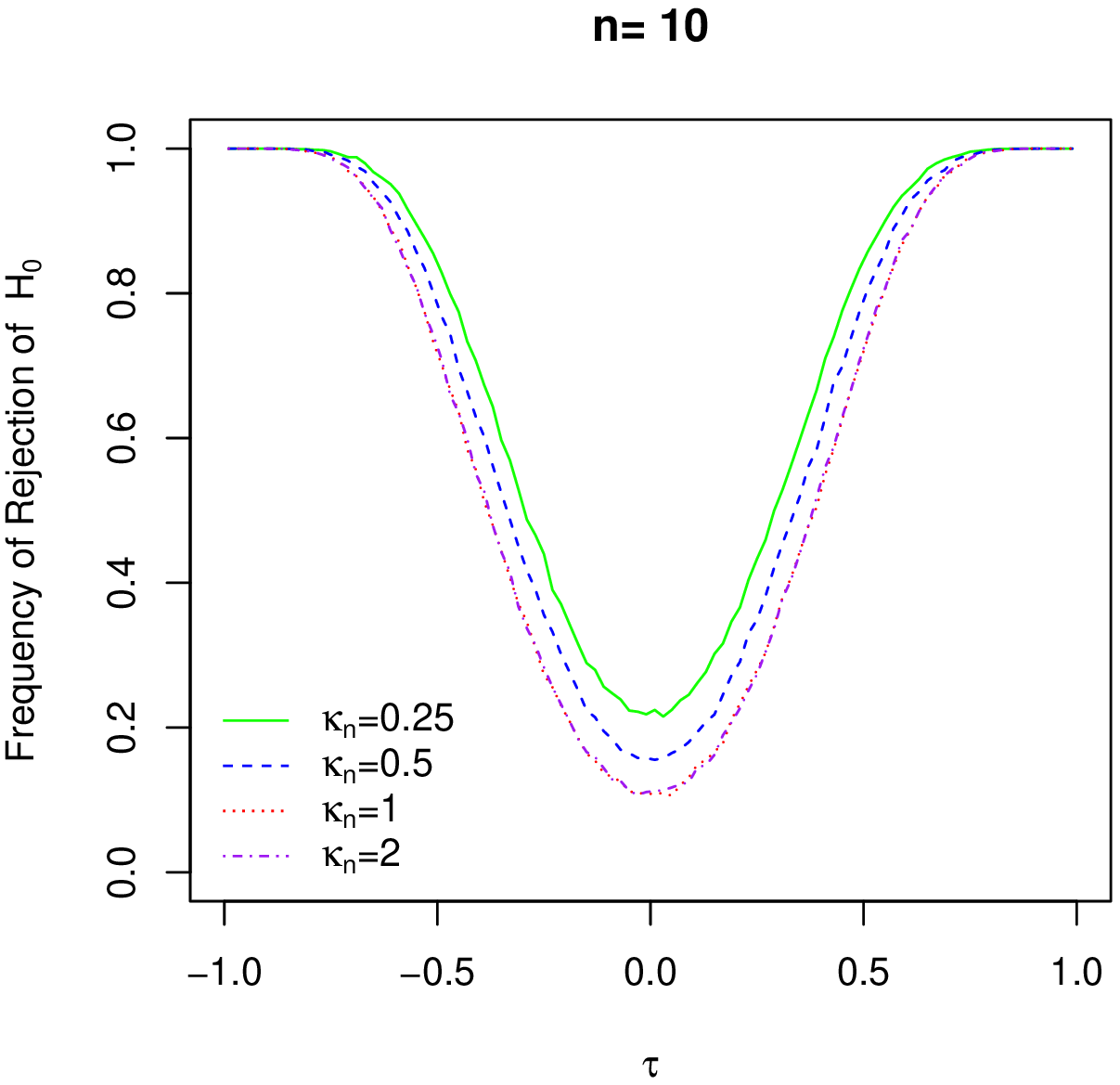}
\includegraphics[width=80mm]{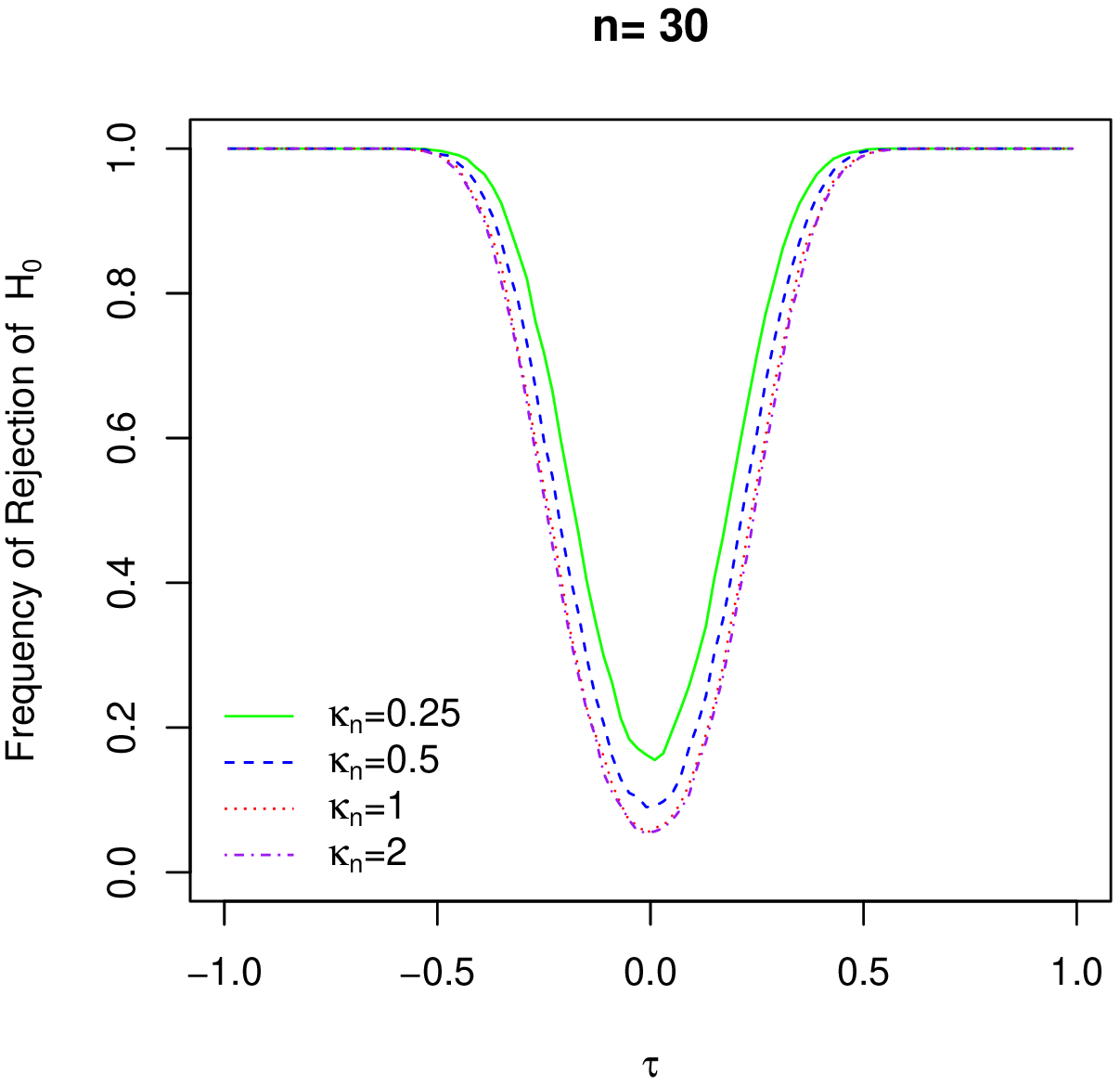}
\includegraphics[width=80mm]{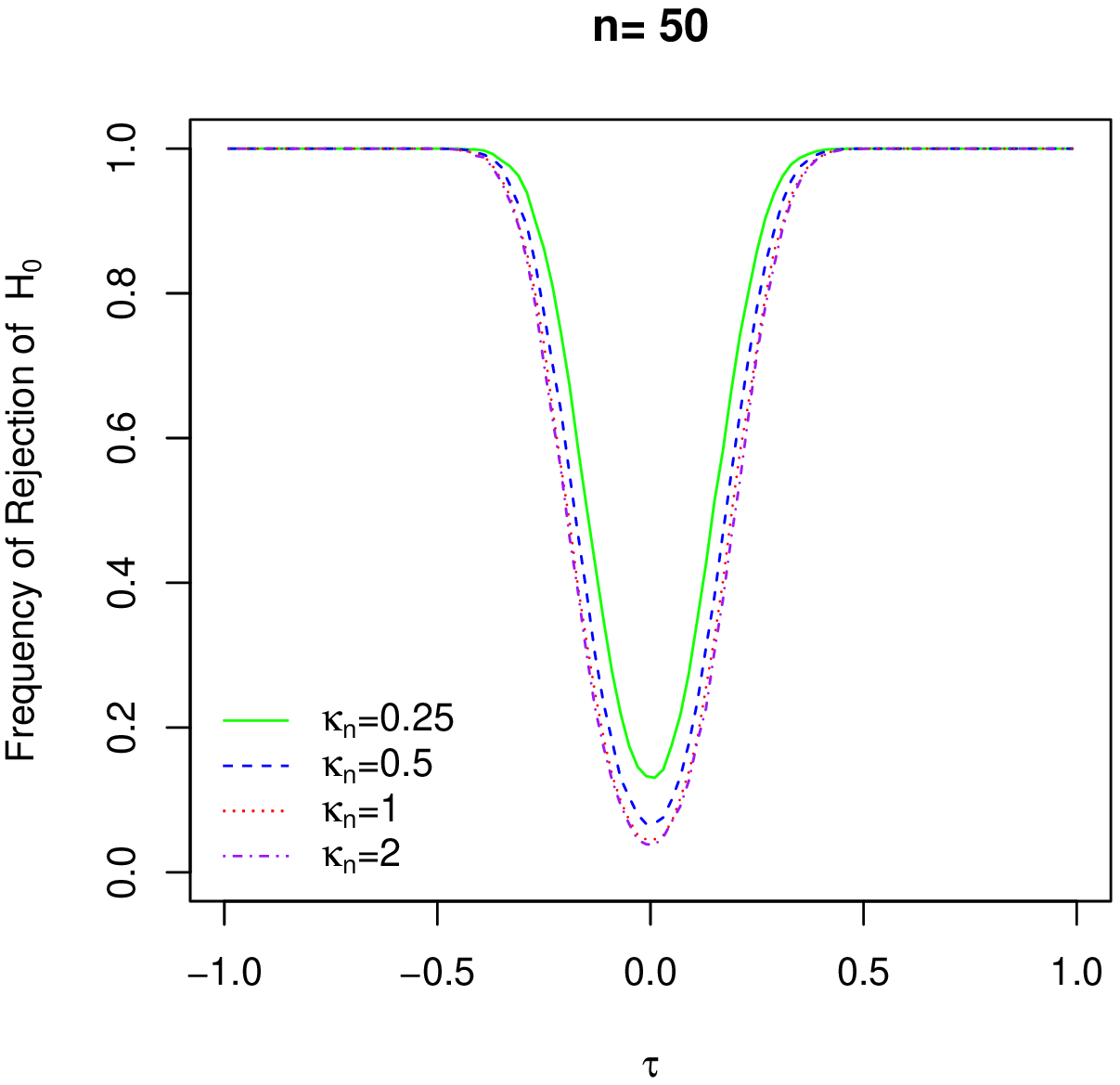}
\includegraphics[width=80mm]{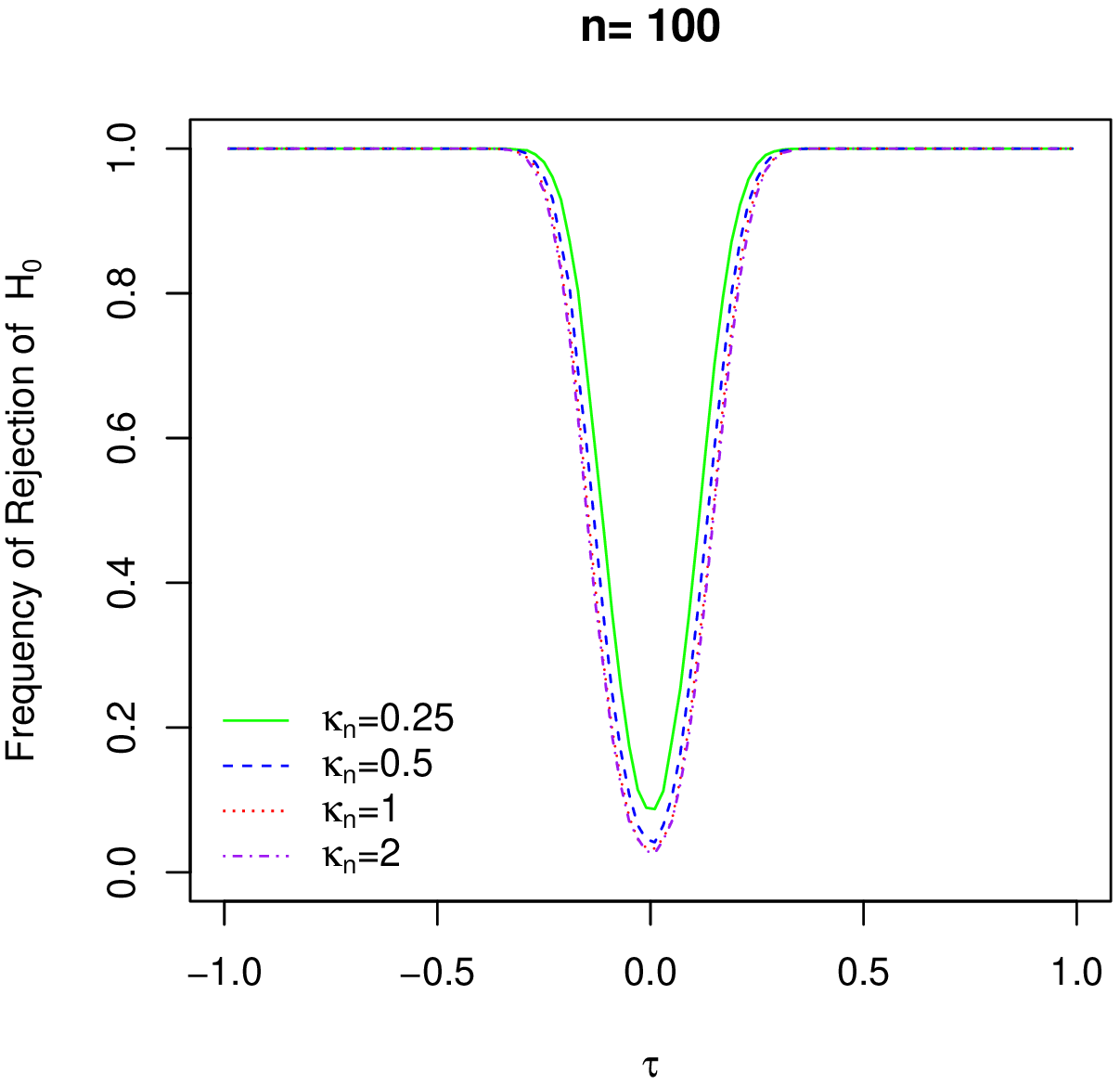}
\caption{The relative frequency of rejection of $H_0$ in (\ref{test:setting:01}) with $\tau_0 = 0$ under BF$_\mathrm{tnorm}$ with $\lambda_n = 0$ and $\kappa_n = \{ 0.25, 0.5, 1, 2\}$.} \label{FIG:FR2}
\end{figure}
}

To further assess the sensitivity analysis of BF$_\mathrm{tnorm}$ with respect to the two hyperparameters $\kappa_n$ and $\lambda_n$, we consider a wide combination of them such that $\lambda_n$ ranging from $-1$ to 1 in increments of 0.01 and $\kappa_n = \{0.25, 0.5, 1, 2\}$ considered above. For this end, we generate another $10,000$ random samples of size $n = 30$ from the bivariate normal copula with the value of $\tau = \{-0.5, 0, 0.5, 0.85\}$. Then, we calculate the BF$_\mathrm{tnorm}$s with each combination of $\kappa_n = \{0.25, 0.5, 1, 2\}$ and $\lambda_n$ ranging from $-1$ to 1 in increments of 0.01 based on the $10,000$ samples. In this way, we obtain the frequency of rejection for each combination of $\lambda_n$, $\kappa_n$, and $\tau$. Numerical results are depicted in Figure \ref{FIG:FR3}. We observe that for a small value of $\kappa_n$ (e.g., $\kappa_n = 0.5$), BF$_\mathrm{tnorm}$ is quite sensitive to the choice of $\lambda_n$, especially when $\lambda_n$ is in the opposite direction of the true value of $\tau$. This phenomenon is quite reasonable, since the probability of the considered prior with a small value of $\kappa_n$ that includes the true value of $\tau$ is almost zero. However, when $\kappa_n \geq 1$, the Bayes factor is quite robust to different values of $\lambda_n$, which is in agreement with our previous sensitivity analysis discussed above. In sum, these simulation studies showed that in the absence of prior knowledge about the unknown parameter $\tau$, we would suggest the use of $(\lambda_n, \kappa_n) = (0, 1)$ for the Bayes factor in (\ref{PBF}) to achieve consistent conclusions, while it is also worth noting that we should try several other values to ensure consistency.

{
\begin{figure}[!t]
\centering
\includegraphics[width=80mm]{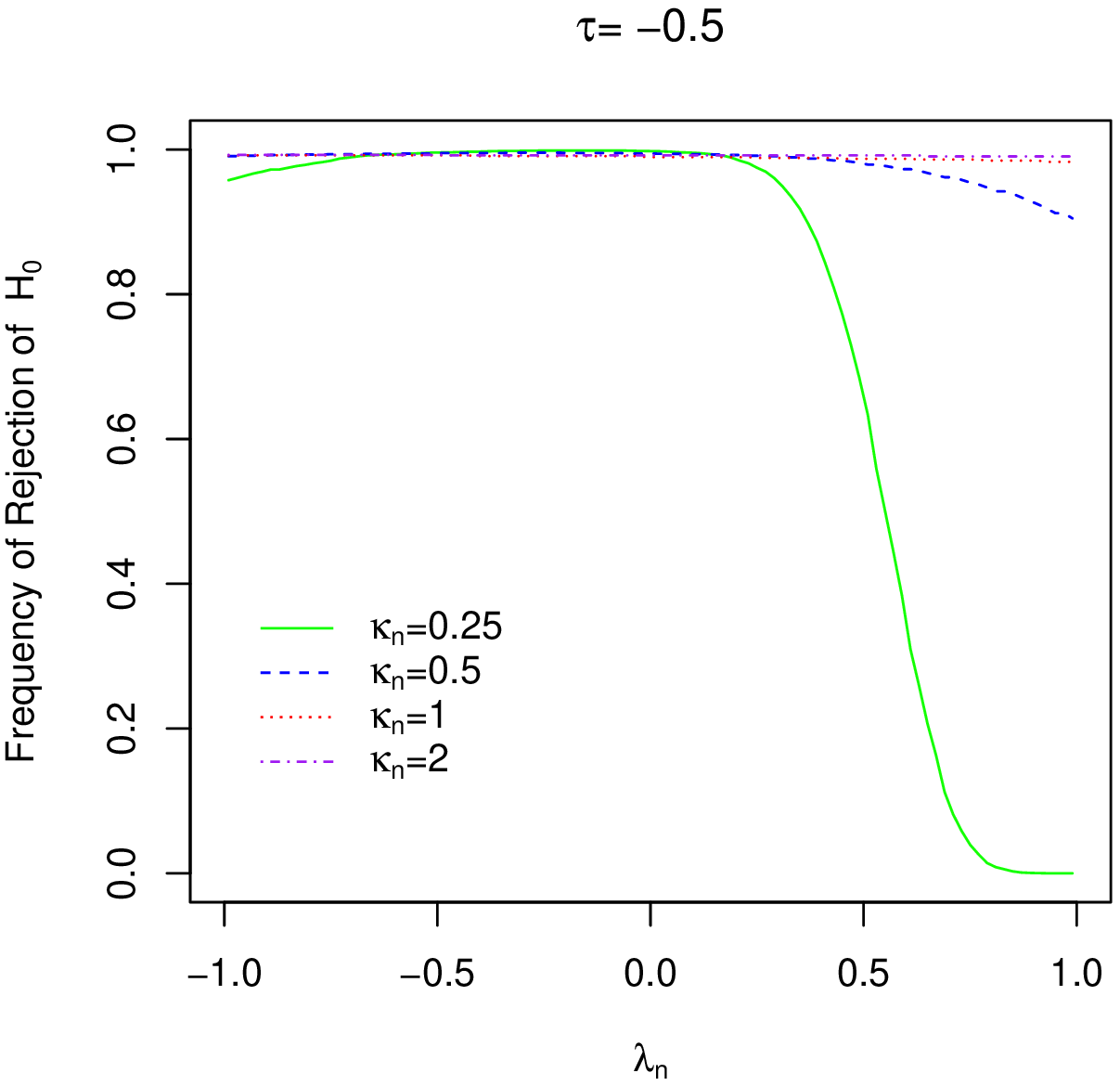}
\includegraphics[width=80mm]{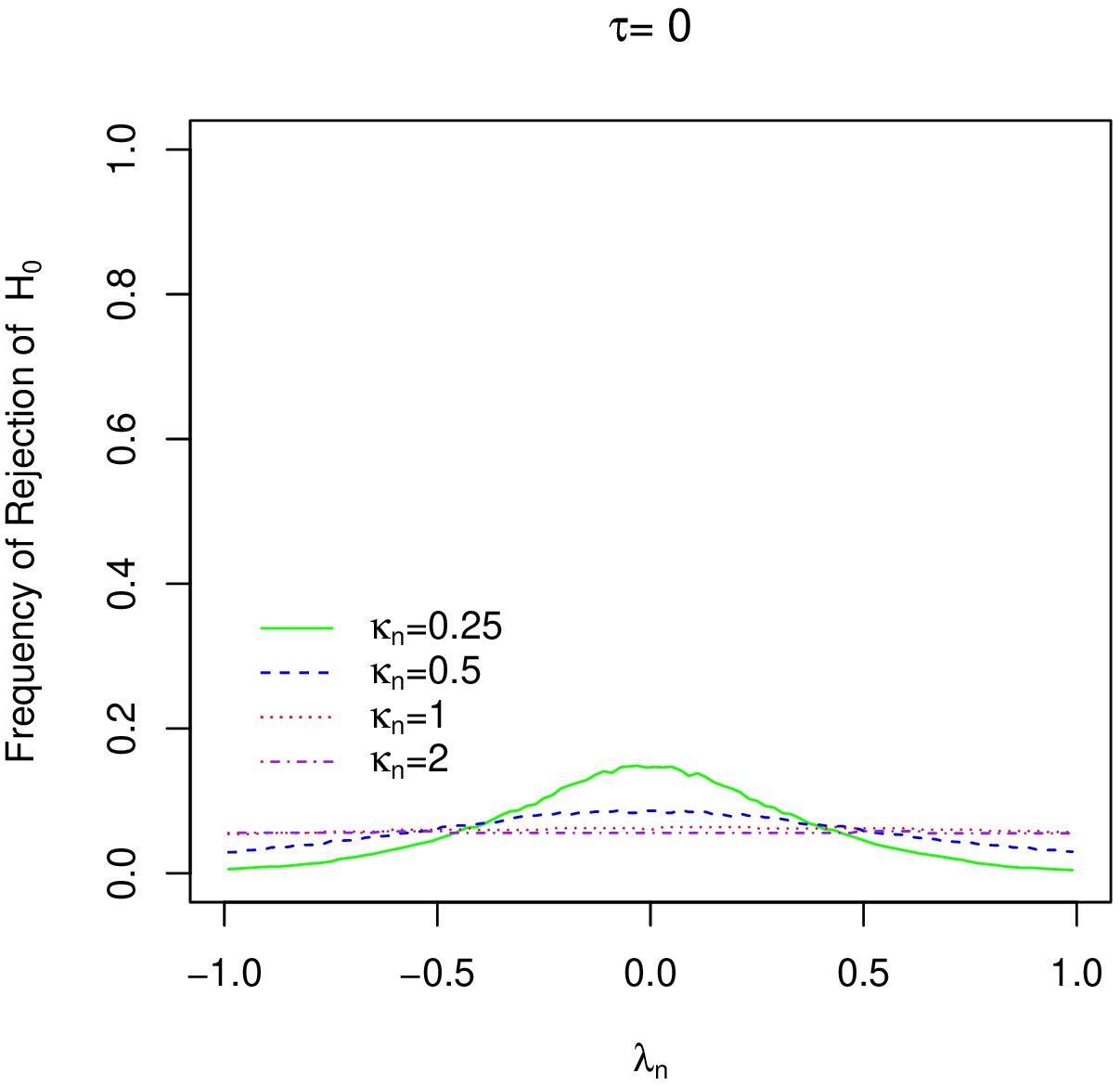}
\includegraphics[width=80mm]{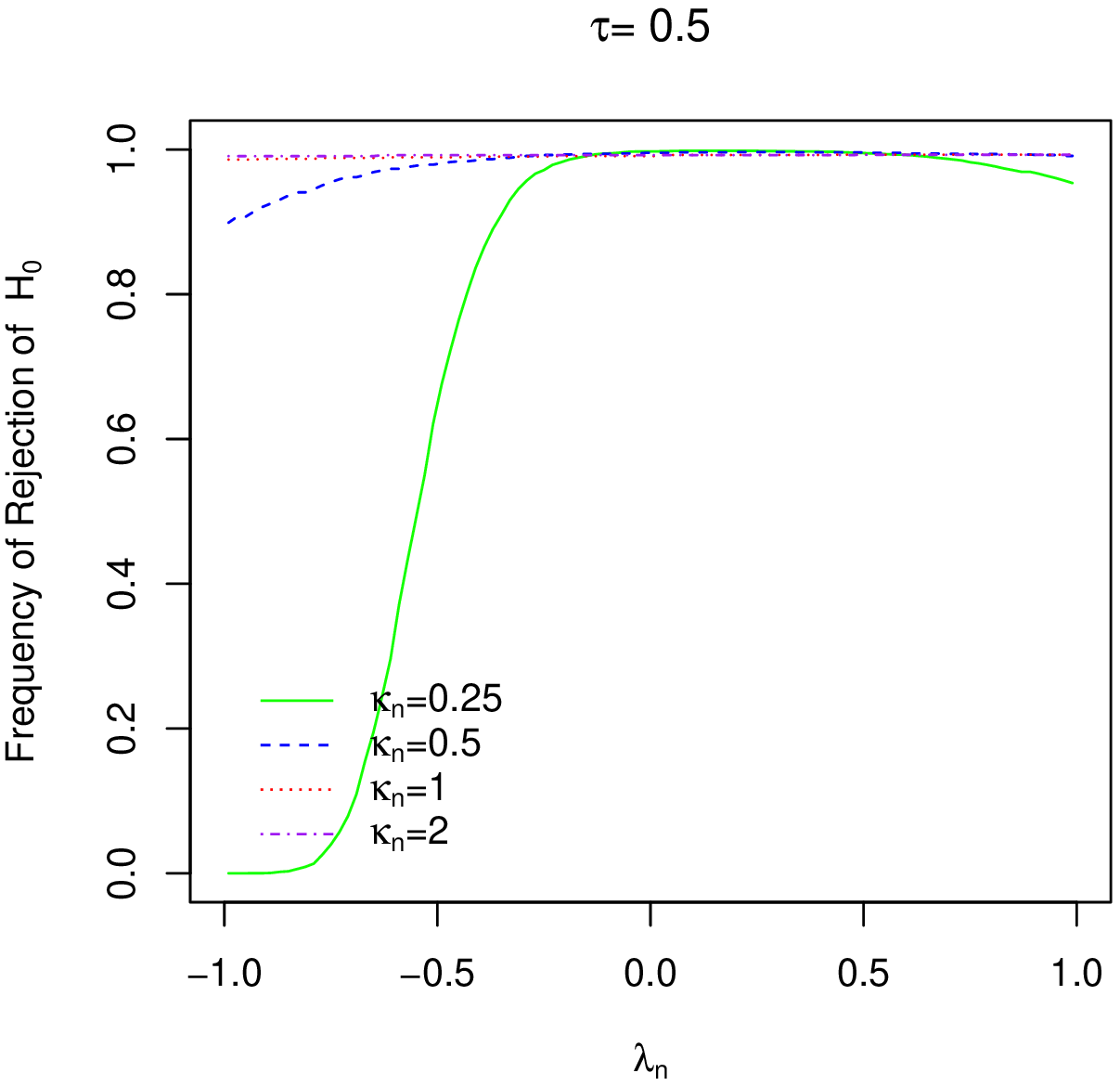}
\includegraphics[width=80mm]{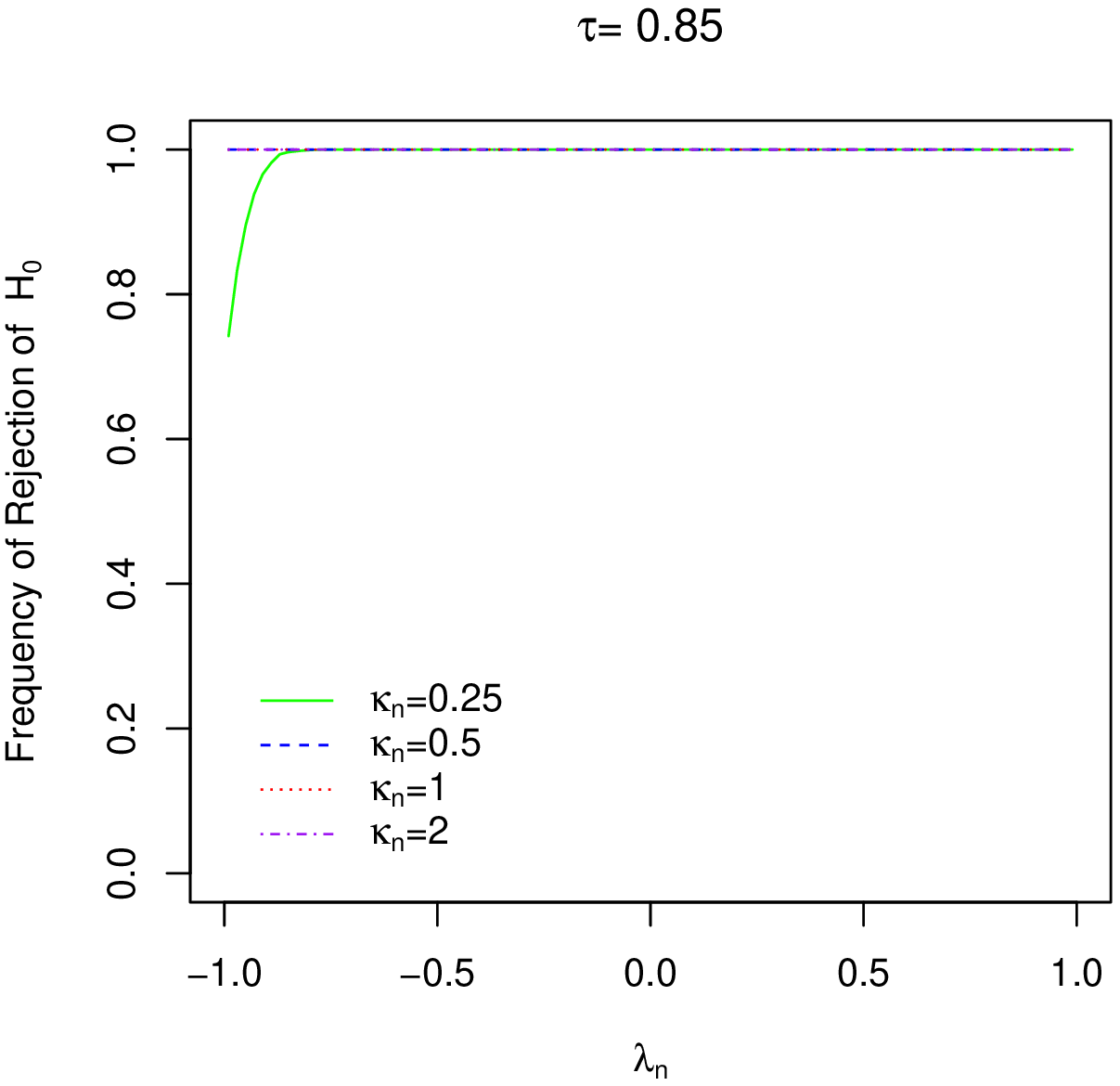}
\caption{The relative frequency of rejection of $H_0$ in (\ref{test:setting:01}) with $\tau_0 = 0$ under BF$_\mathrm{tnorm}$ with $\lambda_n$ ranging from $-1$ to 1 in increments of 0.01 and $\kappa_n = \{ 0.25, 0.5, 1, 2\}$.} \label{FIG:FR3}
\end{figure}
}

\subsection{Simulation 2} \label{simulation:02}

To compare the performance of the Bayes factor BF$_\mathrm{tnorm}$ with $(\lambda_n, \kappa_n) = (0, 1)$ with the default Bayes factor (BF$_\mathrm{default}$) by \cite{Doorn:Ly:Wage:2018}  and the p-value to a significance level $\alpha=0.05$, we follow the same simulation schemes described in Section (\ref{simulation:01}) with the true value of $\tau$ from $-1$ to 1 in increments of 0.1, and the sample size $n= \{10, 30, 50, 100\}$.

Numerical results with different sample sizes are plotted in Figure \ref{FIG:FR1}. Instead of providing exhaustive results for simulation studies, we merely highlight several important findings as follows. (i) When the sample size is small (e.g., $n=10$), the BF$_\mathrm{default}$ test is more likely to be liberal given that its relative frequency of rejecting $H_0$ is far above the significance level from a frequentist perspective. (ii) We observe that for small sample sizes below 30, BF$_\mathrm{tnorm}$ could reach a good compromise between BF$_\mathrm{default}$ and the p-value at the significance level $\alpha=0.05$. (iii) As one expects, when the sample size becomes moderate or large, the three testing procedures behave quite similarly. (iv) The relative frequency of rejecting $H_0$ of BF$_\mathrm{tnorm}$ has a faster decreasing rate to zero than the two testing procedures, validating the consistent behavior of the Bayes factor summarized in Corollary \ref{cor}.

{
\begin{figure}[!t]
\centering
\includegraphics[width=70mm]{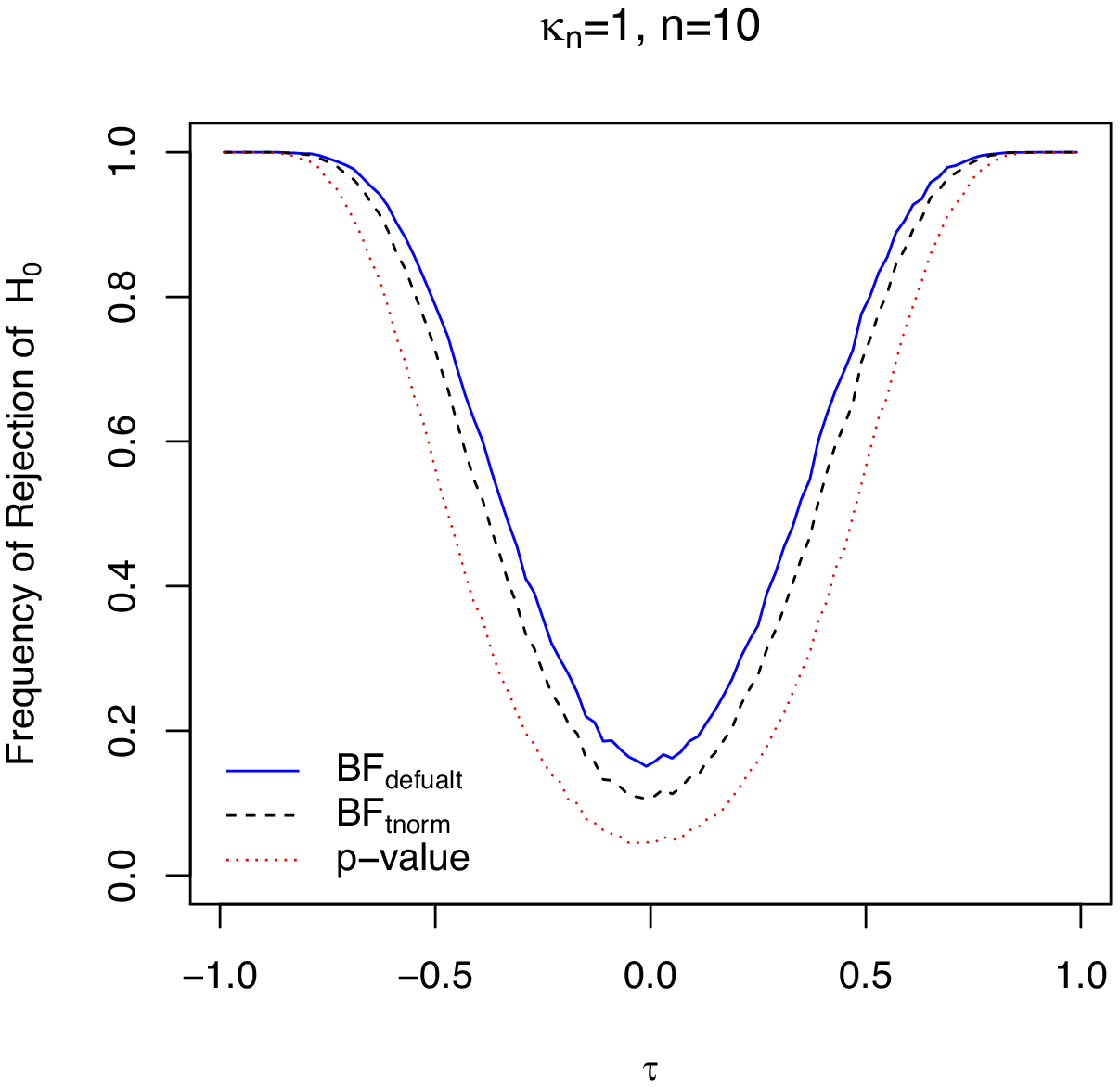}
\includegraphics[width=70mm]{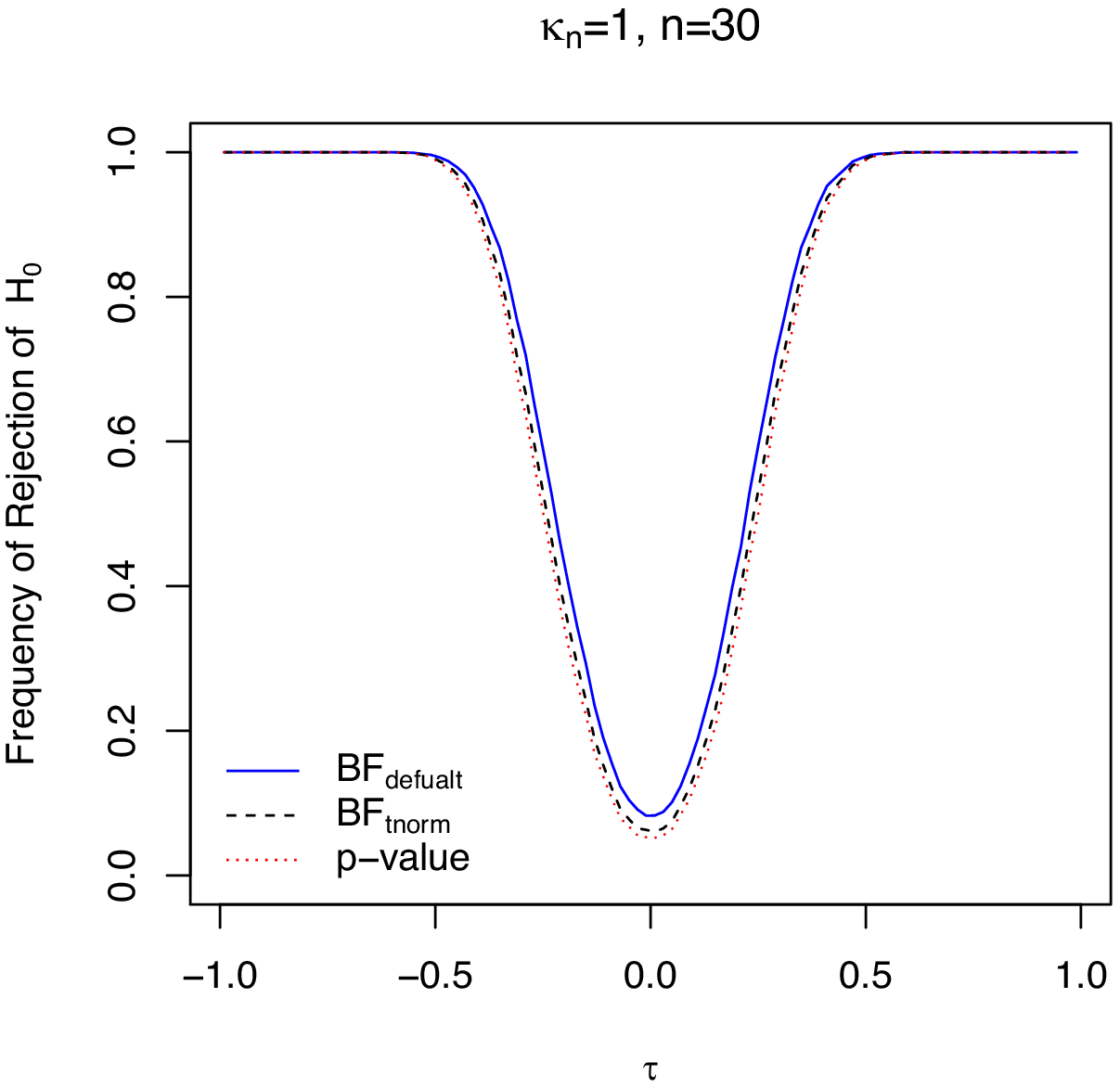}
\includegraphics[width=70mm]{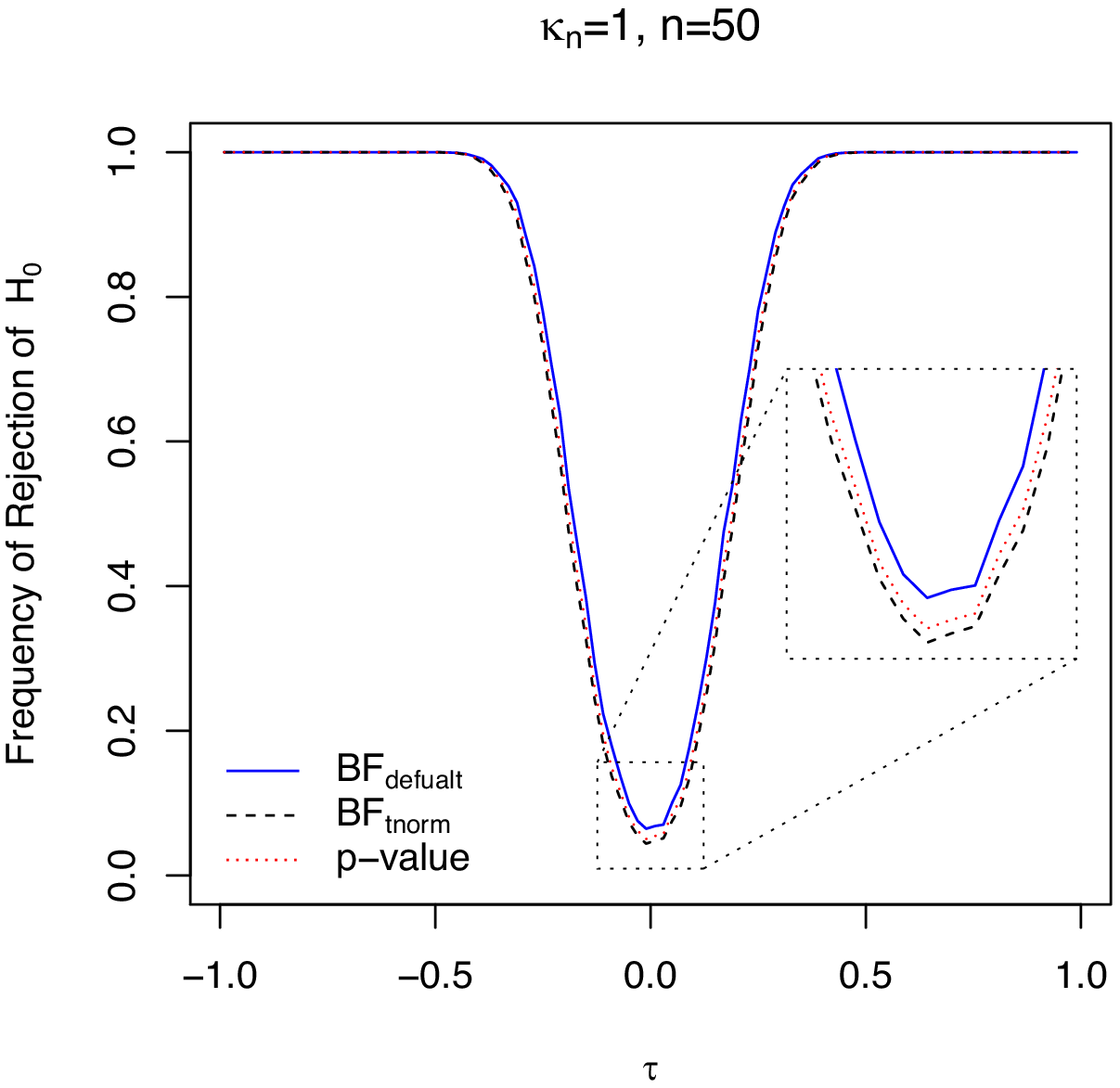}
\includegraphics[width=70mm]{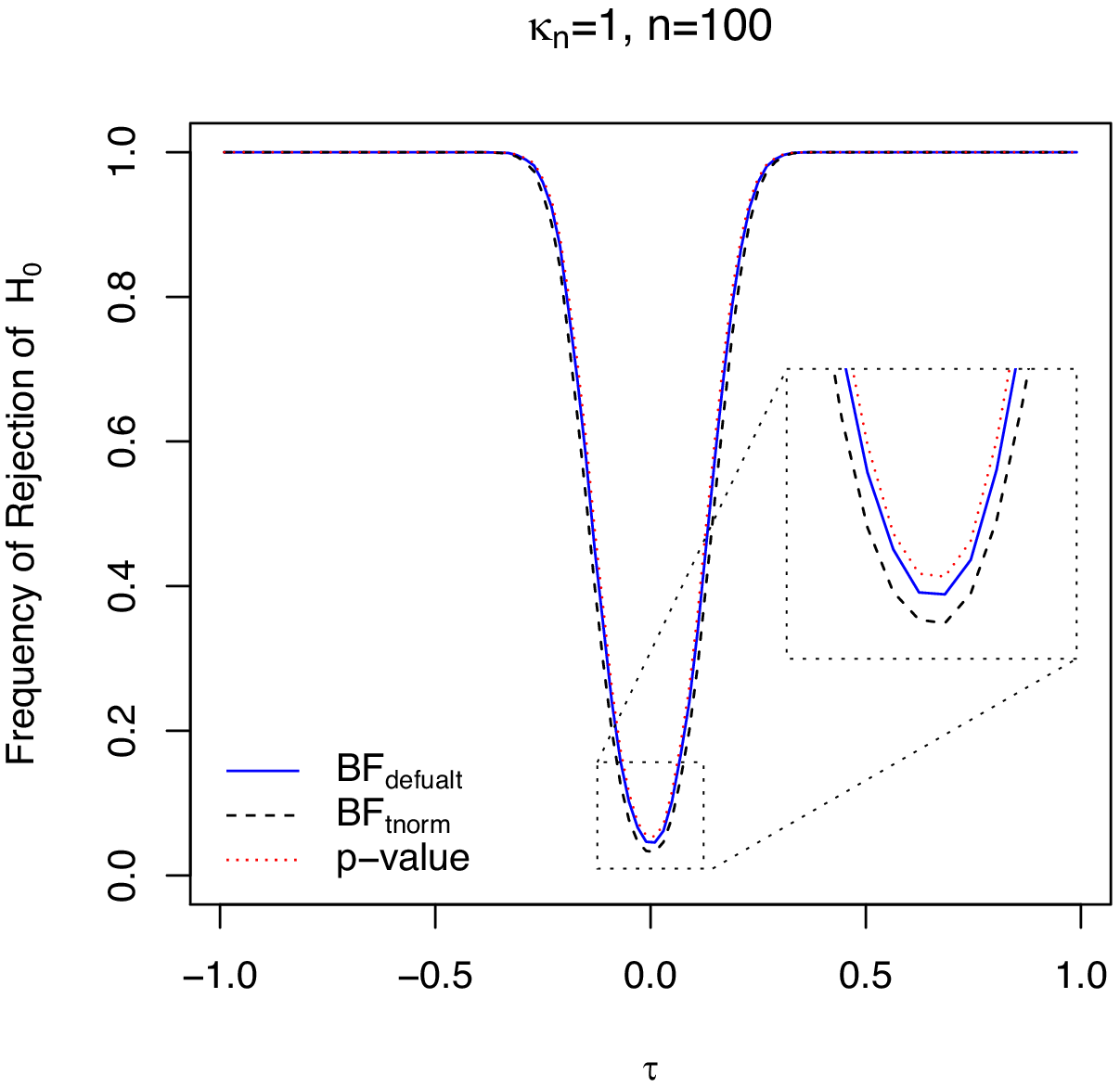}
\caption{The relative frequency of rejection of $H_0$ generated from BF$_\mathrm{tnorm}$ with $(\lambda_n, \kappa_n) = (0, 1)$ and other two testing procedures when $n=\{10, 30, 50, 100\}$.} \label{FIG:FR1}
\end{figure}
}

\section{A real-data application} \label{section:04}

In this section, we illustrate the practical application of the Bayes factor BF$_\mathrm{tnorm}$ through using the data available at \href{https://math.tntech.edu/e-stat/DASL/page11.html}{https://math.tntech.edu/e-stat/DASL/page11.html}. The data set consists of 40 right-handed Anglo introductory psychology students from a large southwestern university. The Magnetic Resonance Imaging (MRI) was adopted to measure the brain size of the participants, and a verbal IQ and a performance IQ score were used to measure the full scale IQ score. We are interested in testing if there exists a correlation between full scale IQ scores and brain size. It can be easily shown that Kendall's  $\tau$ correlation coefficient is 0.3251 with the two-sided p-value of $ p\text{-value} = 0.0035$ The positive value of Kendall's $\tau$ indicates full scale IQ is positively correlated with brain size, and the small p-value shows that the null hypothesis of the independence between full scale IQ and brain size is rejected at the 5\% significance level.

To implement our Bayes factor for this data, we first assume that there is no information about the direction of the relation between full scale IQ scores and brain size, and we may thus assume that $\lambda_n = 0$ to represent prior ignorance. We apply the Bayes factor with different values of $\kappa_n= \{0.25, 0.5, 1, 2\}$. It can be seen from Table \ref{table:bf} that BF$_\mathrm{tnorm}$ is quite robust to the selection of $\kappa_n$ and leads to an identical decision that there is strong evidence supporting the presence of an association between the full scale IQ scores and brain size. For instance, when $(\lambda_n, \kappa_n) = (0, 1)$, the BF$_\mathrm{tnorm}$ is 0.0869 that means the data are $1/0.0869 \approx  11.5$ times more likely to be generated under the alternative hypothesis than under the null hypothesis. Figure \ref{fig:Rplotheatmaprealdata} shows substantial evidence against the null hypothesis for a wide combination of $\lambda_n \in [0, 1]$ and $\kappa_n \in [0, 0.3]$.

{
\begin{table}[h!]
    \centering
    \begin{tabular}{ccccc}
  \hline
 $\kappa_n$ & $0.25 $  & $0.5$ & $1$ &  $2$ \\ \hline
 BF$_{\mathrm{tnorm}}$  & 0.0632  & 0.0708 & 0.0869 & 0.0936\\ \hline  
\end{tabular}
    \caption{The values of BF$_\mathrm{tnorm}$ with $\lambda_n = 0$ and $\kappa_n = \{ 0.25, 0.5, 1, 2\}$.}
    \label{table:bf}
\end{table}

\begin{figure}[!t]
    \centering
    \includegraphics{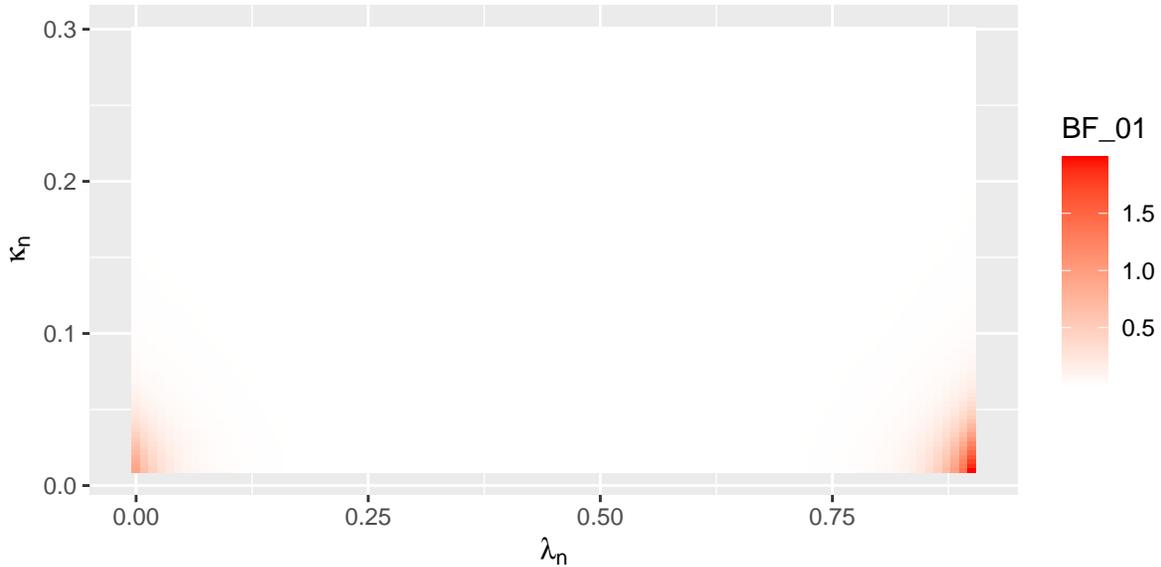}
    \caption{The values of BF$_\mathrm{tnorm}$ with a wide combination of $\lambda_n \in [0, 1]$ and $\kappa_n \in [0, 0.3]$.}
    \label{fig:Rplotheatmaprealdata}
\end{figure}
}

\section{Concluding remarks} \label{section:05}

In this paper, we have proposed an explicit Bayes factor for testing the presence of an association, described by Kendall's $\tau$ coefficient, between the two variables measured on at least an ordinal scale. The proposed Bayes factor enjoys several appealing properties. First, it relies on data only through the Kendall rank correlation coefficient and can thus be easily calculated by practitioners, so long as they are familiar with the Kendall's $\tau$ test. Second, as the sample size approaches infinity, it is consistent whichever hypothesis is true. Third, it can be easily covered in undergraduate and master level statistics course  with an emphasis on Bayesian thinking in nonparametric settings. In addition, numerical results from both simulation studies and a real-data application showed that the performance of the proposed Bayes factor is quite satisfactory and comparable with existing testing procedures in the literature. We hope that the results of this paper will not only facilitate the practical application of Bayesian nonparametric hypothesis testing as an alternative to the classical Kendall's $\tau$ test, but also shed some light on the importance of the hyper-prior specification of the Bayesian methods to students, educators, and researchers.

It is worth noting that owing to the absence of the likelihood functions for the data, we have developed the Bayes factor for the Kendall rank correlation coefficient based on the asymptotic sampling distributions of the test statistic. A natural question to ask, therefore, is that whether these desirable properties of the Bayes factor can be obtained with suitable prior distributions for other nonparametric test statistics having asymptotic normal distributions, which is currently under investigation.  In addition, it may be of interest to consider similar features of the Bayes factors based on nonparametric test statistics having asymptotic $\chi^2$ distributions.

\section*{Acknowledgment}
The authors greatly appreciate the reviewers' thoughtful comments and suggestions, which have significantly improved the quality of the manuscript. This work was based on the first author's dissertation research which was supervised by the corresponding author. The work of Dr. Min Wang was partially supported by the Internal Research Awards (INTRA) program from the UTSA Vice President for Research, Economic Development, and Knowledge Enterprise.

\bibliographystyle{annals}

\section*{Appendix} \label{section:App}

\noindent{\bf Proof of Theorem \ref{theorem:01}}

In order to derive the Bayes factor in (\ref{PBF}), we first need to study asymptotic behaviors of $T^{\ast 2}$ under the null and alternative hypotheses. As the sample size $n$ increases, $T^{\ast 2}$ has a limiting standard normal distribution when $H_0$ is true. Thus, by directly modeling the asymptotic distribution of the test statistic, we obtain the sampling distribution of the data under $H_0$ such that
\begin{equation}\label{eq:ProofTh1}
P(\text{data} \mid H_0) = P(T^{\ast} \mid \tau_0) = \frac{1}{\sqrt{2\pi}}\exp \biggl( -\frac{T^{\ast2}}{2}\biggr),
\end{equation}
since $\Delta = 0$ under $H_0$. However, when $H_1$ is true, $T^{\ast 2}$ has a limiting standard normal distribution with mean $3\sqrt{n}\Delta_n/2$ and variance 1, denoted by $T^\ast \sim \text{N}(3\sqrt{n}\Delta_n/2, 1)$. By specifying the truncated normal distribution as a prior for $\Delta_n$, $\Delta_n\sim TN(\mu_0,\sigma_0,-1-\tau_0,1-\tau_0)$, we have
\begin{eqnarray}\label{eq:ProofTh2}
&&P(\text{data} \mid H_1) = \int P(\text{data} \mid \bftheta_1, H_1)\pi_1(\bftheta_1 \mid  H_1)\, d\bftheta_1,\nonumber\\[3pt] \nonumber
&=&  \int_{-1-\tau_0}^{1-\tau_0}  p(T^\ast \mid \tau_0+\Delta_n)p(\Delta_n)\, d\Delta_n \\
&=&\frac{1}{\Phi\left(\frac{1-\tau_0-\lambda_n}{\kappa_n}\right)-\Phi\left(\frac{-1-\tau_0-\lambda_n}{\kappa_n}\right)}\int_{-1-\tau_0}^{1-\tau_0}\frac{1}{2 \pi\kappa_n}\exp\left\{-\frac{1}{2}\left[\left(T^\ast-\frac{3\sqrt{n}\Delta_n}{2}\right)^2+\frac{1}{\kappa_n^2}\left(\Delta_n-\lambda_n\right)^2\right]\right\}\,d\Delta_n\nonumber\\
&=& \frac{1}{\Phi\left(\frac{1-\tau_0-\lambda_n}{\kappa_n}\right)-\Phi\left(\frac{-1-\tau_0-\lambda_n}{\kappa_n}\right)} \frac{1}{2 \pi\kappa_n} \exp \left\{-\frac{1}{2}\left[- \frac{\left(\frac{T^\ast 3\sqrt{n}}{2}+  \frac{\lambda_n}{\kappa_n^2}\right)^2}{\frac{9n}{4} + \frac{1}{\kappa_n^2} }+ T^{*2}+\frac{\lambda_n^2}{\kappa_n^2} \right]  \right\} \times \nonumber\\[3pt]
&& \int_{-1-\tau_0}^{1-\tau_0} \exp \left\{-\frac{1}{2}\left(\frac{9n}{4} + \frac{1}{\kappa_n^2}\right)\left(\Delta_n-\frac{\frac{T^\ast 3\sqrt{n}}{2}+  \frac{\lambda_n}{\kappa_n^2}}{\frac{9n}{4} + \frac{1}{\kappa_n^2} } \right)^2   \right\}\, d\Delta_n\nonumber\\[3pt]
&=& \frac{\Phi\left(\frac{1-\tau_0-\mu_n}{\sigma_n}\right)-\Phi\left(\frac{-1-\tau_0-\mu_n}{\sigma_n}\right)}{\Phi\left(\frac{1-\tau_0-\lambda_n}{\kappa_n}\right)-\Phi\left(\frac{-1-\tau_0-\lambda_n}{\kappa_n}\right)}
 \frac{\sigma_n}{\sqrt{2 \pi}\kappa_n} \exp \left\{-\frac{1}{2}\left[- \frac{\left(\frac{T^\ast 3\sqrt{n}}{2}+  \frac{\lambda_n}{\kappa_n^2}\right)^2}{\frac{9n}{4} + \frac{1}{\kappa_n^2} }+ T^{*2}+\frac{\lambda_n^2}{\kappa_n^2} \right]  \right\},
\end{eqnarray}
where $\mu_n$ and $\sigma_n$ are given in (\ref{eq:postMeanVar}). Therefore, the Bayes factor in (\ref{BF:00}) is given by
\begin{eqnarray*}
\mathrm{BF}_{01} &=&  \frac{P(T^{\ast2} \mid \tau_0)}{\int_{-1-\tau_0}^{1-\tau_0}  p(T^\ast \mid \tau_0+\Delta_n)p(\Delta_n)\, d\Delta_n}\\[3pt]
&=&\sqrt{\frac{9n\kappa_n^2}{4}+1} ~\exp \left\{\frac{1}{2}\left(\frac{\lambda_n^2}{\kappa_n^2}- \frac{\mu_n^2 }{\sigma_n^2} \right)  \right\}\cdot \frac{\Phi\left(\frac{1-\tau_0-\mu_n}{\sigma_n}\right)-\Phi\left(\frac{-1-\tau_0-\mu_n}{\sigma_n}\right)}{\Phi\left(\frac{1-\tau_0-\lambda_n}{\kappa_n}\right)-\Phi\left(\frac{-1-\tau_0-\lambda_n}{\kappa_n}\right)}.
\end{eqnarray*}
This completed the proof of Theorem \ref{theorem:01}.

\vskip4mm
\noindent{\bf Proof of Theorem \ref{theorem:02}}

For $\lambda_n = O(n^{-a})$ and $\kappa_n = O(n^{-b})$ for $0\le a\le b <{1}/{2}$, we consider each term of the proposed Bayes factor in (\ref{PBF}) as follows.
\begin{enumerate}
    \item[(a)] $\sqrt{\frac{9n\kappa_n^2}{4}+1} = \sqrt{O(n^{1-2b})} = O(n^{1/2-b}).$ In addition, we observe that $\sigma_n^{-1} = O(n^{1/2})$.
    \item[(b)] $\frac{\pm 1-\tau_0-\lambda_n}{\kappa_n} = (\pm 1- \tau_0)O(n^b) - O(n^{b-a})\rightarrow \pm\infty$ as $b>0$, or two different constants as $b=a=0$. Hence  $0<\Phi\bigl(\frac{1-\tau_0-\lambda_n}{\kappa_n}\bigr)-\Phi\bigl(\frac{-1-\tau_0-\lambda_n}{\kappa_n}\bigr)\le 1$.
    \item[(c)] For  $\Phi\bigl(\frac{1-\tau_0-\mu_n}{\sigma_n}\bigr)-\Phi\bigl(\frac{-1-\tau_0-\mu_n}{\sigma_n}\bigr)$, we consider \begin{align*}
       \frac{\pm 1 -\tau_0 - \mu_n}{\sigma_n} &= \frac{(\pm 1-\tau_0)\kappa_n^{-1}\left(\frac{9n\kappa_n^2}{4}+1\right) -\frac{3}{2}\sqrt{n}\kappa_n T^* -\frac{\lambda_n}{\kappa_n}}{\sqrt{\frac{9n\kappa_n^2}{4}+1}} \\
       &=\begin{cases}
       O(n^{b-1/2)}\left[(\pm 1 -\tau_0)O(n^{1-b}) -O(n^{1/2-b}) -O(n^{b-a})\right],& \text{when $H_0$ is true},\\\frac{\frac{9n\kappa_n}{4}\left(\pm 1 -\tau_0-\Delta_n\right) +(\pm 1 -\tau_0)\kappa_n^{-1}-\frac{3}{2}\sqrt{n}\kappa_n\left(T^* - \frac{3}{2}\sqrt{n}\Delta_n\right)-\frac{\lambda_n}{\kappa_n}}{\sqrt{\frac{9n\kappa_n^2}{4}+1}},&\text{when $H_1$ is true},
       \end{cases}\\
       &= \begin{cases}
       (\pm 1 -\tau_0)O(n^{1/2}) + O(1),& H_0,\\(\pm 1 -\tau_0 -\Delta_n)O(n^{1/2}) +(\pm 1-\tau_0)O(n^{2b-1/2}) +O(1) + O(n^{2b-a-1/2}),&H_1.
       \end{cases}
    \end{align*}
When $H_0$ is true, we clearly observe that as $n$ approaches infinity,  $(\pm 1 -\tau_0)O(n^{1/2})\rightarrow\pm \infty$. On the other hand, since $\Delta_n$ is between $-1-\tau_0$ and $1-\tau_0$, $\pm 1-\tau_0 -\Delta_n$ have opposite signs. When $H_1$ is true, with $b-a\ge 0$ and $b<\frac{1}{2}$, $\frac{\pm 1 -\tau_0 - \mu_n}{\sigma_n}\rightarrow\infty$ as $n\rightarrow\infty$. Therefore, we conclude that $\Phi\bigl(\frac{1-\tau_0-\mu_n}{\sigma_n}\bigr)-\Phi\bigl(\frac{-1-\tau_0-\mu_n}{\sigma_n}\bigr)\rightarrow 1$ as $n$ tends to infinity.
    \item[(d)] We now consider
    \begin{align*}
       \frac{\lambda_n^2}{\kappa_n^2}- \frac{\mu_n^2}{\sigma_n^2} &= \frac{\lambda_n^2}{\kappa_n^2} - \frac{\left(\frac{3}{2}\sqrt{n}\kappa_n T^*-\frac{ \lambda_n}{\kappa_n}\right)^2}{\frac{9}{4}n\kappa_n^2 +1} =\frac{\frac{9}{4}n\lambda_n^2 -\frac{9}{4}n\kappa_n^2T^{*2} +3\sqrt{n}\lambda_n T^*}{\frac{9}{4}n\kappa_n^2+1}\\
       &= \begin{cases}
           O(n^{2b-1})\left[O(n^{1-2a}) +O(n^{1-2b}) +O(n^{1/2-a})\right],&\text{when $H_0$ is true}, \\\frac{\frac{9}{4}n\lambda_n^2 -\frac{9}{4}n\kappa_n^2\left(T^{*} -\frac{3}{2}\sqrt{n}\Delta_n +\frac{3}{2}\sqrt{n} \Delta_n\right)^2 +3\sqrt{n}\lambda_n \left(T^* -\frac{3}{2}\sqrt{n}\Delta_n +\frac{3}{2}\sqrt{n} \Delta_n \right)}{\frac{9}{4}n\kappa_n^2+1},& \text{when $H_1$ is true},
       \end{cases}\\
       &= \begin{cases}
           O(n^{2(b-a)})+ O(n^{2b-a-1/2}),&\text{when $H_0$ is true}, \\O(n^{2(b-a)}) + O(n^{2b-a-1/2}) -O(n),& \text{when $H_1$ is true},
    \end{cases}
    \end{align*}
    where the coefficient in $O(n)$ for $H_1$ case is positive.  Since $b-a\ge 0$, the power of $n$ when $H_0$ is true is at least non-negative. On the other hand, since $b<1/2$, the first term in $\mathrm{BF}_{01}$ of (\ref{PBF}) always goes to infinity. Hence, $\mathrm{BF}_{01}\rightarrow\infty$ as $n\rightarrow\infty$ when $H_0$ is true. On the other hand, when $H_1$ is true, since both $2(b-a)<1$ and $2b-a-1/2<1$, the last term of the above expression is dominated by $-O(n)$. Hence, we can see that  $\exp\left[\frac{1}{2}\left( \frac{\lambda_n^2}{\kappa_n^2}- \frac{\mu_n^2}{\sigma_n^2}\right)\right]\rightarrow0$ as $n\rightarrow\infty$. So is $\mathrm{BF}_{01}$ in (\ref{PBF}). We conclude that our Bayes factor in (\ref{PBF}) is consistent whichever the true hypothesis is. This completed the proof of Theorem \ref{theorem:02}.
    \end{enumerate}

\end{document}